# Sociological cycles: the accumulated discrepancy between appearance and reality as driver

Joachim Maier

Max-Planck-Institute for Solid State Research, Heisenbergstr.1, D-70569 Stuttgart, Germany,

s.weiglein@fkf.mpg.de

Oscillations are observed in all branches of science and culture, ranging from the behavior of elementary particles, atoms, molecules in simple chemical or physical systems or even in complex organisms, up to oscillations of the behavior of complex organisms such as human beings[1]. Examples of the latter are phenomena as different as waves of public taste, particularly obvious in fashion, or periodically repeated surplus or deficiency of qualified jobs in a certain profession. In the focus of the presentation are those cycles that are characterized by periodical sequences of over- and under-estimation (“hype cycles”). They are not only ubiquitous but also of great impact on and hence of great interest for society. Many of the other oscillation modes, though, can be mapped on this.

If ensembles of simple atomic or molecular particles are considered, there are essentially two approaches to treat such phenomena, (i) thermodynamic considerations that involve relations between fluxes (reaction rates) and driving forces, and (ii) kinetic considerations that are concerned with reaction rates in terms of collision events (encounter events, usually treated by master-equations). These approaches generally assume so-called “mean-field” concepts (a better term would be: homogenization concepts) in which the averaged overall property of an ensemble of constituents is replaced by the property of a virtual uniform ensemble of representative (average) constituents.

Needless to say, that such simple relations could never describe human behavior in most of its respects. The proof is simple: Through reflexion a human being is always able by a purposefully imposed feedback loop to counteract a prognosticated behavior. But in particular, if phenomena of large groups are concerned, there can be cases in which a highly individual behavior may not be very relevant[2-7], and a common mass behavior may be approached by mean-field concepts. Consider as example growth models of populations[8-11] or models of collective behavior of crowds in the case of a panic[6]. Further simple examples for the usefulness of kinetic equations are mass action problems such as the "restaurant problem" that considers the occupation of seats in a restaurant[12] or prey-predator models treated by reaction- or diffusion-controlled Lotka-Volterra kinetics[9]. The latter has become a standard model to describe oscillatory behavior in non-equilibrium systems. There it is non-linearity connected with a sufficiently complex mechanism that includes autocatalytic elements, which causes oscillations.

What, however, makes sociological problems very different from simple atomistic ones is the fact that memory effects play an important role. Here it is not so that encounter processes lead ad-hoc to defined results rather decisions are based on the judgment of an actual situation on the basis of experience (prehistory). The oscillating phenomena that we are considering here are indeed enabled by memory effects but can still be tackled in a mean-field sense.

It is noteworthy that it is memory here and not complexity of the mechanism which leads to oscillations under otherwise very weak assumptions. In order to exemplify the effect of memory we consider, as a preparation, a simple collision model, that we later even directly apply to describe fashion cycles (see also supplementary information 1). It refers to a first step of our consideration but is already very revealing in this context. The results are shown by Figure 1.

In this collision model a blue particle whenever meeting a catalyst transforms, in the memory-less case, into a red one and vice versa. Starting with a non-equilibrium distribution, in the memory-less case the expected exponential relaxation takes place. Now, a memory effect can be included by the rule that a blue (red) particle has to meet the catalyst not only once, but for n-times. As obvious from the figure such introduction of memory leads to a damped-oscillatory behavior which is the less damped the greater *n*. (Note that the evaluation of the logistic curve leads to similar responses if time-delay is introduced[13-15].)

From the stand-point of collision kinetics the described memory-less reaction is a pseudo-first order reaction showing an exponential time-behavior. Naturally such a behavior can be recast in terms of a proportionality between reaction rate and deviation of actual value from equilibrium value as driving force (see suppl. 1). In the case with memory we refer to a sequence of pseudo-first order reactions as far as the individual reaction partner is concerned. In order now to describe processes being more general than encounter processes but exhibiting memory effects, we will thus make the "thermodynamic" ansatz that the reaction rate is proportional to the time integral of the deviation from actual value and "equilibrium" value.

A relevant master example (later we will address further examples) may be the introduction of an innovation that had initially been oversold. After having realized that the expectation does not match reality, the interest decreases and—owing to deep disappointment—more strongly than the innovation deserves. This frustration phase in which the innovation is undervalued may trigger the next overestimation phase.

This pattern is met in various fields: in science (consider number of publications in initially hot topics, e.g. superconductivity), in technology (a most recent example: initial down-playing of the importance of electro-mobility in traffic now may lead to a hype in this respect) but also in politics (the sudden break-down of a political systems causes overestimation of advantages of the competing system, or to give another example, an overrated newly elected political leader may soon perceive frustration effects in the next elections). Fashion cycles can also be subsumed un-der this pattern: If we assume as major driving force the tendency to be exclusive, a certain fashion, once having become popular, shows the tendency to be getting out-dated until the memory of it has faded away.

The processes to be described here are such that the estimated value of a given property (i.e. the perceived or expected value, $E$) deviates strongly from its real, true value ($e$). After the constraints responsible for the initial valuation are removed, the system is prone to oscillate.

In realistic situations the oscillations may become highly damped or may already after one cycle reach a plateau (cf. Gartner hype cycle[16]), what can be ascribed by a superposition of irreversible processes, e.g. by learning effects[17].

In this contribution we are primarily interested in the underlying aptitude of a sociological system to oscillate and emphasize that this is the consequence of typical human behavior (connected with memory, experience, and inclination of being persistent) that endows the system with the necessary inertia, without which a spontaneous response would have to be expected. In the beginning, thus, we will abstract from effects that lead to damping or other realistic additional reactions.

The model to be outlined will be kept as simple as possible and be characterized by a few sociologically important parameters.

Even though our main interest is devoted to the oscillation phase, we will later shortly deal with the initial phase setting the initial condition, not only as this prehistory determines the amplitudes of the oscillations but also since important sociological scenarios such as dictatorship in terms of exclusivity of power or of information as well as tabooization (caused by external or internal mechanisms) can be easily addressed in this way. In our context such a pre-phase is formally most simply modeled by considering the evolution of the sociological parameters into stationary values.

While such parameter variations could also account for learning effects and hence for realistic cycles, the obvious isomorphicity of the sociological model to be described with oscillations in physical processes (free oscillations of a spring or an electrical circuit consisting of capacity and inductivity, after having removed the bias) suggests damping to be modeled by introducing a direct response channel into the *sociological equivalent circuit*.

Let us first describe our model. In spite of similarities it uses ingredients that makes it different from usual thermodynamic descriptions: (i) It considers as “driving function” ($\Delta$) for the changes in sociological activity ($a$) the difference between *expectation value* ($E$) and *realistic value* ($e$), or in general but less precise terms the discrepancy between *seeming* and *being*. (ii) It considers the *time integral* over this “driving function” (and not $\Delta$ directly) as being proportional to the time changes of the activity.

## The “hype-problem” and the role of accumulated discrepancy between expectation and reality

For clarity’s sake, but without significant loss of generality, let us consider the following master example. A new technology is appraised and a certain, initially oversold success proclaimed ($E$) with the real success ($e$) remaining behind $E$ by a factor $\varepsilon^{-1} = e/E < 1$. The difference $\Delta \equiv E - e = \left(1 - 1/\varepsilon\right)E$ may be called here the exaggeration gap. The promised success will prompt new activities contributing to the success of the technology. This activity ($a$) may be e.g. measured by the number of persons involved in this business. The fact that it is $\int (E-e)dt$, i.e. the accumulated discrepancy between expectation (or more generally apparent value) and reality (real value), that drives the reaction rate ($\dot{a}$) in many real problems, agrees with daily life experi-ence. It is not the immediate deviation between $E$ and $e$ that will trigger changes in the activity. But if this discrepancy is obvious over a certain time interval, this will happen.

Furthermore, it is reasonable as first approximation to take the activity $a$ as proportional to the promised success *E*, through $\eta = a/E$. (Examples may be the number of enterprises, of employers, of publications dealing with a topic of promised success *E*.) If not outer constraints dominate, the activity will be the more pronounced the greater the expected success. In fact in many realistic examples (see below) *E* will just be measured by *a* so that $\eta$ might simply be unity.

These three assumptions naturally lead to (*t*: time, dot: time derivative):

$$\dot{a} = -\alpha \int_0^t (E-e)dt + \dot{a}(t=0)$$

or for constant prefactor $\alpha$ (1)

$$\ddot{a} = -\alpha(E-e) = -\frac{\alpha}{\eta}(1-\frac{1}{\varepsilon})a \equiv -\omega^2 a$$

If, the parameters $\varepsilon$ and $\eta$ are indeed constants, we arrive at the harmonic oscillation problem with oscillations around *a* = 0, as shown in Fig. 2 ($\omega$: angular frequency). The interpretation is as follows: A gap between expected success and real success and consequently an over-activity characterize the over-motivation phase. This gap does not directly affect the rate $\dot{a}$ but rather the acceleration $\ddot{a}$. Only after this gap has continually increased for a while, the activity will decrease. In other words, there is a certain credit that needs to be exhausted. This then leads to the pendulum swinging into the other direction. The over-motivation is replaced by frustration and now the success is judged too negatively. The real success is comparatively larger but it takes again some time until this is counter-acted. It is worth stating again that such a simple situation would not lead to oscillations if $\dot{a}$ were proportional to (*E* – *e*) directly, rather *a* would—depending on the sign of (*E* – *e*)—exponentially decay or continuously increase. It is obviously the described inertia that is necessary for the addressed oscillations.

In more general terms these parameters need to be defined differentially. If we draw the analogue to the electrical situation, $-d\ddot{a}/d\Delta = \alpha$ corresponds to an inverse social inductance (L) whereas $da/d\Delta = \eta\varepsilon/(1-\varepsilon)$ corresponds to a social capacitance (C). Both elements are connected in parallel and hence refer to the same driving function $\Delta$. If the parameters are constant, as assumed in this contribution ($C \equiv a/\Delta = \eta \frac{\varepsilon}{\varepsilon - 1}$ and $L \equiv 1/\alpha$), oscillations are observed with the angular frequency $\omega = 1/LC$.

So far the system oscillates around $a = E = e = 0$ with the consequence that *a*, *E*, *e* become negative. It is clear that generally these quantities have to be understood as deviations from the "equilibrium state". If $e$, $E$ did denote absolute (in the sense of total) values, that are only allowed to be slightly modified by the oscillation, non-zero values of $e$ and $E$ could not at the same time be equal and proportional through a non-zero $\varepsilon$. Rather in the linear approximation the relation between the absolute values need to include an additive constant. If for a moment we use the above symbols $E_t$, $e_t$ for the absolute values rather than for the perturbations, then $E_t = \varepsilon e_t + const$. The additional degree of freedom in terms of the constant may be used to define the equilibrium state. It is certainly a reasonable procedure to characterize the equilibrium by the postulate that $E_t$ and $e_t$ become equal, i.e. by the postulate that the discrepancy between being and seeming disappears ($\widehat{E}_t = \widehat{e}_t$, if the arc denotes the equilibrium case). This leads to $const = (1-\varepsilon)\widehat{e}$ and allows us to term the equilibrium "*expectation equilibrium*". Using this definition, $E_t$ and $e_t$ refer to the same equilibrium value and we can employ Eq. (1) that only considers their difference, also for the absolute values. (Note that this equilibrium state is not an equilibrium state in the thermodynamic sense.) A special but important case refers to the situa-tion, where $e_t$ stays constant, i.e. keeps it “equilibrium value”. In other words $e = 0$ and $\varepsilon$ is formally infinite. We address this point below.

More complex mechanisms (e.g. of Lotka-Volterra type) can easily introduce phase shifts between $a$ and $\Delta$, or weighting functions may be introduced by replacing the integration by a convolution[18,19]. We will return to the latter point at the end. At the moment, however, we want to keep the description as simple as possible.

As with any model, the information with respect to specific situations is hidden in the parameters, here $\varepsilon$, $\eta$ and $\alpha$. Before we inspect them, let us investigate as to what conclusions other, in particular non-linear $e(a)$ functionalities would lead. (This means naturally that the above formally introduced $\varepsilon$ is no longer a meaningful parameter.) For this purpose let us drop the index $t$ and understand $e, E, a$ as absolute values; furthermore, for simplicity’s sake, we set the remaining proportionality factors to unity and additive constants to zero.

The simplest case is a constant success value. An example where this may be approximately fulfilled is a rather saturated technology in the case of which the real success is not sensitively altered by varying activities. This can, however, be very different as far as the expectation value is concerned. A pertinent example is a politically sensitive technology (e.g. energy technology) hence giving rise to possibly great gaps between promise and reality. In this case we get oscillations of $E$ and $a$ but not for $e$ which is always given by an “equilibrium value”.

A related class of problems that can be treated in that way are the fashion problems (cf. suppl. 1). The popularity of given names is one of the few examples where reliable statistics are available. The ranking of given names can be followed for Germany for the entire last century. Figure 3 shows two examples, a male and a female name the popularity of which initially decreased presumably owing to excessive use, and then experienced a revival. The mapping of this problem on our master example is evident if we identify $E$ (or $a$ which we do not need to distinguish here) with the popularity of the name measured in terms of its actual frequency, i.e. taking account of the sociological context (“apparent” or “perceived value”), while $e = \hat{e}$ is a measure of how appropriate the name is irrespective of its frequency (“true value”). For simplicity we concentrate on the exclusivity effect as the major driver: If the name has become overpopular there is a tendency not to use the name. Owing to the memory this continues to be the case initially even for $E$ being smaller than *e*, where after it is the accumulated rareness of the name that increases its use again. The half-period of about 50 years is a typical period within which an out-dated given name got sufficiently out of “sight”.

It is very revealing to investigate powers in $\Delta \propto a^n - a$ other than $n = 0$ or $n = 1$. Here it may suffice to make only a few remarks. If $0 < n < 1$, $a^n$ exceeds $a$ for $a < 1$; then $a$ increases in an accelerated way allowing for one-sided oscillations (see supplementary information Supplementary 2), i.e. the inflection point around which the function may oscillate is generated on the positive side. For $a > 1$, $a^n$ is smaller than $a$, and is increasingly so for greater $t$, leading to deceleration and non-monotonicity. For $n > 1$ the situation is inverse, first characterized by a deceleration and then—if $a$ exceeds unity—by ever increasing acceleration. The latter then leads to a catastrophic behavior.

Typical solutions for $n = 2,3$ are based on tangent or cotangent functions; they show a periodical but not harmonically oscillating behavior (see supplementary information 2).

The diverging behavior contained in these solutions makes the full time evolution unrealistic but highlights the explosivity of such a situation. (The term ‘unrealistic” implies that in such situations the parameters will certainly not remain invariant). At any rate, the behavior in the first period is intriguing. The steep increase of *a*, *E*, *e* near the asymptotes means that the impact of this technology steeply grows as may be the case for a very successful young technology, which may easily outcompete others.

An interesting question that may arise here is how sociological systems (or technologies) compete with each other. There is an adequate kinetic model of how molecular species compete about common limited resources[20], which led to an epitomization of Darwin's theory. Setting up a similar sociological theory by implementing memory effects as in the above would be a very intriguing challenge but beyond the scope of this paper (see also supplementary information 3).

## Model parameters

Let us discuss the meaning of the above model parameters. In the above defined terms α is a measure of how rapidly the activity is accelerated given a certain integral discrepancy between promise and reality ($1/\alpha$: "degree of inertia"), it is a measure of reaction readiness, short-windedness of the system and is proportional to the square of the frequency ($\omega$) of the oscillations $\cong \sqrt{\alpha\left(1-\frac{1}{\varepsilon}\right)\frac{1}{\eta}}$. Low frequencies indicate a sluggish reaction, the actors are rather inert. Very high frequencies indicate a hysterical behavior. Examples can range from quick responses by hectic persons to collective cultural movements that involve generation changes.

Also the other model parameters $\eta$ and $\varepsilon$ are contained in $\omega$. As in many cases *E* is simply measured in terms of *a*, the meaning of $\eta$ is trivial. It simply is then a conversion or sensitivity factor. In the above example the $\eta$-factor characterizes the motivation and determines how strongly the promised success impacts on activity ("motivability").

Finally, $\varepsilon$, i.e. the ratio *E/e*, is a measure of the gap ($E - e$) per real success ($= \varepsilon - 1$), and hence characterizes the hype most directly. It determines how great the gap is at given *e* and may be called "exaggeration factor". If this factor is close to unity, one deals with extremely careful and conservative estimates of the situation ($E \simeq e$); $\varepsilon = 1$ characterizes a completely realistic judgement and conservation of a stable expectation “equilibrium”. As the above considerations have shown, $\varepsilon$ need not be constant to enable oscillations but constancy is, as far as its role in Eq. (1) is concerned, sufficient. Constancy of $\varepsilon$ is best fulfilled for small excitations, i.e. for not too large a discrepancy concerning expectation and reality. In cases that the real success value is fixed (i.e. $e$ conceived as deviation from equilibrium value is zero) but $E$ changes, $\varepsilon$ is formally infinitely large.

As we take all these parameters time-invariant, in our simple model an important qualitative characterization is the following “aphorism”: *Whatever is responsible for the pendulum swinging too much to the right, is also responsible for its extreme swinging to the other direction.*

In the same sense then: One does not do any good on a long run for propagating an idea or a concept if one is overstressing its advantage; the almost unavoidable backslash will punish the exaggeration. Except the fact that probably a certain $\Delta$ is needed to overcome initial “nucleation barriers”, one is well advised to stay as close as possible to the “expectation equilibrium”.

How great the amplitudes of the oscillations are, is determined by the prehistory (cf. bias of a mechanical or electrical oscillatory system). A complete theory that includes prehistory would explain the evolution of the harmonic oscillations by the inner dynamics of the overall system. As this has to involve specific information on the specific case, such an attempt would be beyond the scope of this paper. A few remarks may suffice.

In concrete or abstract terms the initial situation is dictated by the constraining forces (analogous to the force that extends a spring). A relevant pre-phase may consist in the introduction of an innovation. In such a pre-phase $\Delta$ may be dictated by the exclusivity of information that e.g. a company possesses when announcing an innovation before the true value can be tested.

Here let us take a purely heuristic route and let us come back to the question of initial conditions by considering the physical master example of a vibrating spring which is characterized by the isomorphic differential equation. The frequency is fixed by the force constant and mass of the spring, but the amplitude determined by the initial condition ($\sqrt{a^2(t=0)+\dot{a}^2(t=0)/\omega^2}$ [21]). Such an initial condition can be characterized by a force compensating the spring force that is suddenly released but also by a sudden variation in the spring constant (or the mass) in the pre-phase**Error! Bookmark not defined.**.

A simple case of a pre-phase setting the initial conditions for *a* is a phase in which the activity is suppressed or kept at a low level for political reasons (dictatorship) or for reasons of "political correctness" (taboo-phase) [5,11]. After overcoming this phase the system undergoes free oscillations with an amplitude set by this phase. Another example may be the deliberate down-playing of the success expectation. (As already mentioned, a recent example is the down-playing of the significance of electromobility in the past as not to endanger the established combustion technology in electrotraction).

Unlike materials objects met in oscillating circuits such as springs, coils, capacitors etc., in sociological systems the parameters ($\alpha, \eta, \varepsilon$) are sociologically determined. In the framework of our simple model we thus characterize the pre-phases by artificial settings of these parameters followed by a relaxation towards stationary values (now termed $\alpha_\infty, \eta_\infty, \varepsilon_\infty$ etc.). Let us briefly investigate two simple cases, with opposite effects on the development of frequency and amplitude. One example may be that $\alpha$ is artificially set to zero (or the "degree of inertia" $1/\alpha$, i.e. the social inductance to infinity) and hence, even if a gap between $E$ and $e$, is realized, a counteraction ( as in the case of an infinite mass) is impeded. This could be directly related with political issues, e.g. that in a totalitarian system the discrepancy between propaganda and real value may be perceived over a while but no collective reaction is allowed to occur (another example is that a product is announced and oversold, yet counter-acting measures cannot be undertaken before the product is on the free market). After suspension of the dictatorship-phase $\alpha$ may develop into the above stationary value, now called $\alpha_\infty$ with a time constant $\tau_\alpha$, according to

$$\alpha(t) = \alpha_\infty\left(1-e^{-t/\tau_\alpha}\right)$$

The second example is that in spite of a finite $\Delta$, the activity is remaining on a low level owing to an artificially small value of $C \equiv \eta\frac{\varepsilon}{\varepsilon-1}$. The sociological reason may be that in view of the sociological context people do not see the perspective to change the situation. In the last case we may similarly formulate $C = C_\infty\left(1-e^{-t/\tau_C}\right)$.

If we adopt such simplified time evolutions, we obtain the differential equation

$\ddot{a} = -\frac{\alpha_\infty\left(1-e^{-t/\tau_\alpha}\right)}{C_\infty\left(1-e^{-t/\tau_C}\right)}a.$ In many cases the two time constants may be sufficiently different, so that the treatment can be decoupled. Assuming arbitrarily $\tau_C \ll \tau_\alpha$ the long-time solution based on Bessel functions and sketched in Fig. 4, is met.

The Bessel functions yield weakly damped harmonic oscillations in the very long time limit (with a damping factor $\propto t^{-1/2}$). More details are given in supplementary 4a .If $\tau_C \gg \tau_\alpha$ the solutions are given by hypergeometric series (see supplementary 4b) having related long-time properties.

An explicit sociological model would allow for a more accurate description of the above phase transitions and a more exact time evolution of the parameters. Breakdown of what is termed taboo-phase in Fig. 3 can happen extrinsically (e.g. by external force) but also intrinsically. Typical kinetic phase transition models[11,22] may involve fluctuations (around the dictated values) leading to instabilities by self-amplifications and finally to the collapse of the initial situation (cf. also supplementary 3).

Such specific considerations are beyond the scope of the present contribution.

Let us now discuss an example in greater detail for which “experimental results” are available. It refers to the popularity of painters in the last century as measured by “google-culturomics”. Michel et al.[23] compare the citations of Marc Chagall in the UK and in Germany in the 20th century. The data are given in Fig. 5. They attributed the difference to the suppression of this artist’s appreciation by the dictatorship during World War II in Germany. Apart from a slight overall increase both curves show oscillations with a period slightly less than 10 years. This may be understood as typical fashion cycles (see suppl. 1). While the normalized behavior for the UK is rather stationary, in Germany, after World War II heavily damped oscillations are observed that develop towards the behavior in the UK. For a rough description we can consider *e* (realistic value) as being given by the normalized mean value. Then, a description according to the behavior just discussed is possible (cf. Fig. 3), where it is the overcoming of the depression phase that leads to oscillatory transients in the Germany example which are very close indeed to the behavior discussed in supplementary 4a $\left(\alpha=\alpha_\infty\left(1-e^{-t/\tau_\alpha}\right)\right)$.

**Damped curves, realistic cycles and equivalent circuits**

In realistic sociological systems undamped oscillations are not met at least not for many cycles. We mentioned initially Gartner’s hype cycle**Error! Bookmark not defined.** which shows that usually after a phase of enthusiasm, a frustration phase occurs which soon can lead to a plateau, named *expectation equilibrium* ($E = e$) in our model. Such a sudden equilibration can be reached by learning effects that have not been considered.

This again can be heuristically implemented by time dependent $\varepsilon$, $\eta$ and $\alpha$ parameters what would then change amplitude (damping) and frequency. In supplementary 5a the effect of an ex-ponentially decaying motivability ŋ is investigated, leading to $\ddot{a} = -const\ e^{+t/\tau'} a$ .

Similar results are obtained by testing a hyperbolic decay (supplementary 5b). The results yield already very realistic shapes.

There is also another simple approach, guided by the analogy with the oscillating circuits in physics, wherein we can mimic damping by admixing a direct response term ($\dot{a} \propto E - e$) complementing the sociological equivalent circuit to form a parallel L-C-R circuit (see suppl. 6). The magnitude of the prefactor would in electrical language be proportional to a conductance ($R^{-1}$). This value or better the ratio between this parameter and $\alpha$ decides upon the quickness of the response. This "direct" spontaneous social mechanism acts as a valve or leakage channel mitigating the initial discrepancy between seeming and being. If the latter is absent there is no damping; if it completely dominates, the situation will relax towards the expectation equilibrium without oscillation. As the accumulative response channel can be characterized by a convolution

$$\dot{a}(t) - \dot{a}(0) \propto \int_0^t f(\tau - t)\Delta(\tau)d\tau \qquad (2)$$

with $f \equiv 1$ (I-element in the language of system-theory), and the direct channel by the convolution with $f = \delta$-function, (P-element) one could formulate the general situation by a convolution $f * \Delta$ with $f$ revealing P-I characteristics. In the language of system theory we refer to admittance (impedance) that is characterized by resistive and inductive influences. (The capacitive relation $a \propto \Delta$ corresponds to a D-characteristic if $\dot{a}$ is taken as out-put signal.)

Even though still heuristic, such approaches seem straightforward for explaining a highly damped curve such as the Gartner curve, and appear more natural than conceiving it as an artificial composite of two mechanisms with different time constants, viz. of a bell-shaped enthusiasm/disappointment response (i.e. as a curve that is ascribed to a quick emotional response) and a more sustainable logical s-shaped response as done in literature***Error! Bookmark not defined.***. Here we would generally describe it as a mixture of direct and accumulative reactions, e.g. reflecting different mechanisms or differently acting social sub-groups (see Fig. 6), or alternatively by an implementation of learning effects (learning from the previous phase) into an otherwise oscillating rigid system. It is worth emphasizing that learning effects (in contrast to the simple damped oscillation) can also take account of a time dependence of the frequency (supplementary 5).

The simple L-C model that is in the foreground of this contribution does not include irreversibility. (This feature was introduced above by the learning effects through the parameters.) In the L-C-R model the irreversibility was introduced by admixing resistive effects.

Just the opposite path is taken by the third approach that was already touched upon in the very beginning, where we used collision kinetics. This collision kinetics is per se irreversible (R-C elements); an oscillatory behavior is arrived at by introducing a memory through the trick that collisions have to take place several times in order to be efficient. A pertinent example is fashion cycles. Let us consider a carrier of blue cloth who on meeting another carrier of blue cloth is stimulated to change to the alternative red color ("exclusivity effect"). Neglecting other "reactions" (such as blue converting to red when meeting red corresponding to an infection effect) we simply can formulate

$$B + B \to R + R. \qquad (3)$$

Given the validity of collision kinetics, such a simple mechanism can never lead to oscillations. Rather starting with blue only leads to a monotonic relaxation towards the equilibrium which is—for otherwise equivalent conditions concerning *B* and *R*—characterized by equipartition. Yet introduction of a memory changes the picture. If we assume that *B* converts to *R* only after having met *B* n-times (see suppl. 1) damped oscillations towards the equilibrium distribution occur, with the damping being the less pronounced the greater n.

With these considerations of initial conditions and more realistic behavioral aspects we did not intend to distract from the major point namely highlighting the general tendency of many sociological systems to oscillate even under simple mechanistic conditions, rather it was meant to show the aptitude of the simple model for refinement.

In short, we presented a simple but realistic model of over-/underestimation cycles based on the accumulated discrepancy between perceived value and real value, or in more general terms between seeming and being. Moreover the model will be of use (with appropriate renaming of the variables and parameters) whenever effects of deviations from the true value (in the most general sense of the word) are of sociological influence. What makes sociological systems easily prone to oscillate (even for simple mechanisms) is the significance of memory and persistence (rather than of spontaneous decisions) that endows the system with the necessary inertia. The model is purposefully strongly simplified but believed to hit the basic points and hence to enable sharper definitions of otherwise extremely loosely defined terms. (It is not in conflict with the fact that sociological events can be mechanistically complex and highly non-linear.) The model stresses the applicability of sociological equivalent circuits and also contains the flexibility to incorporate the sociological prehistory as well as learning effects leading to a more realistic description.

**Figures**

**Figure 1**

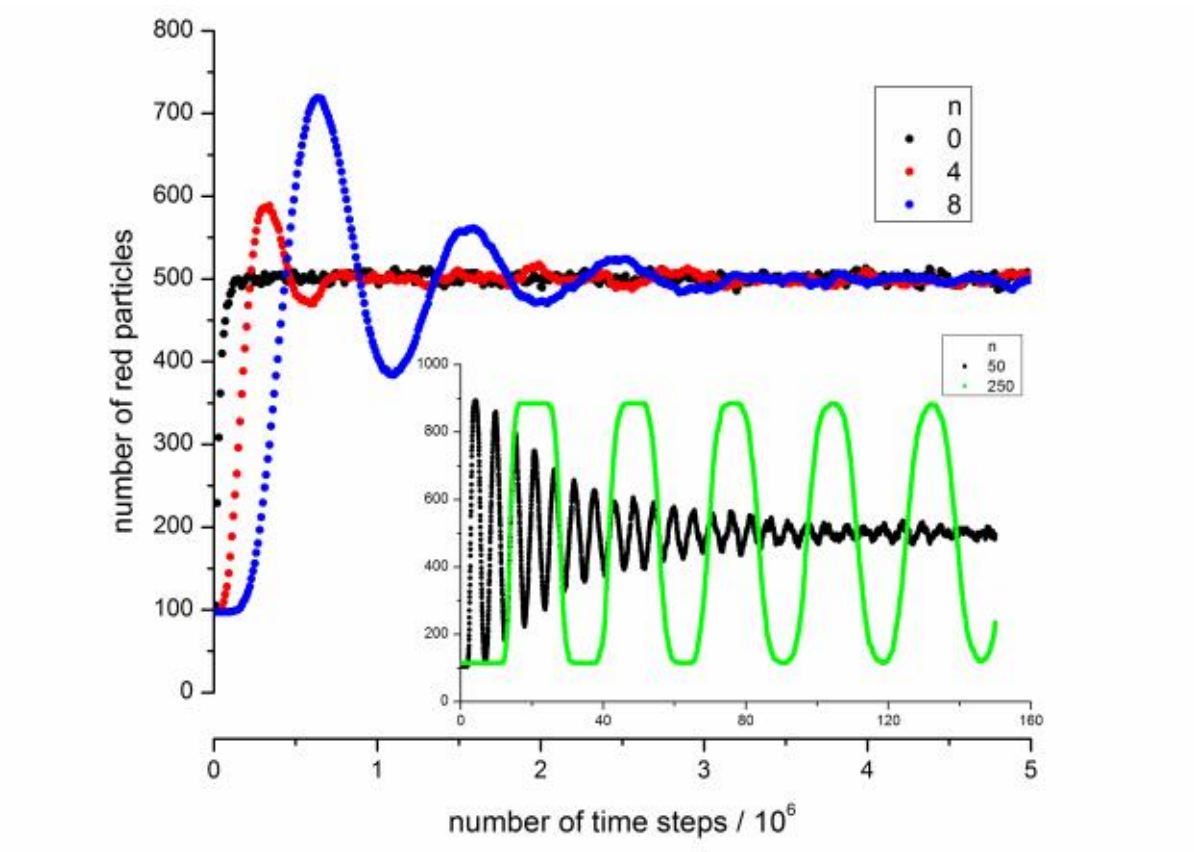

Monte-Carlo collision experiments with memory (n). Unlike the memory-less ♦ memory-free???♦ case (n = 0) for which one observes an exponential relaxation, the memory effect leads to a damped-oscillatory-behavior. The model assumes that a blue particle which has met a catalyst for the n-th time changes into a red one vice versa. (Initially: 900 blue particles, 100 red particles. The number of catalysts is 9000. Cf. also supplementary 1.

**Figure 2**

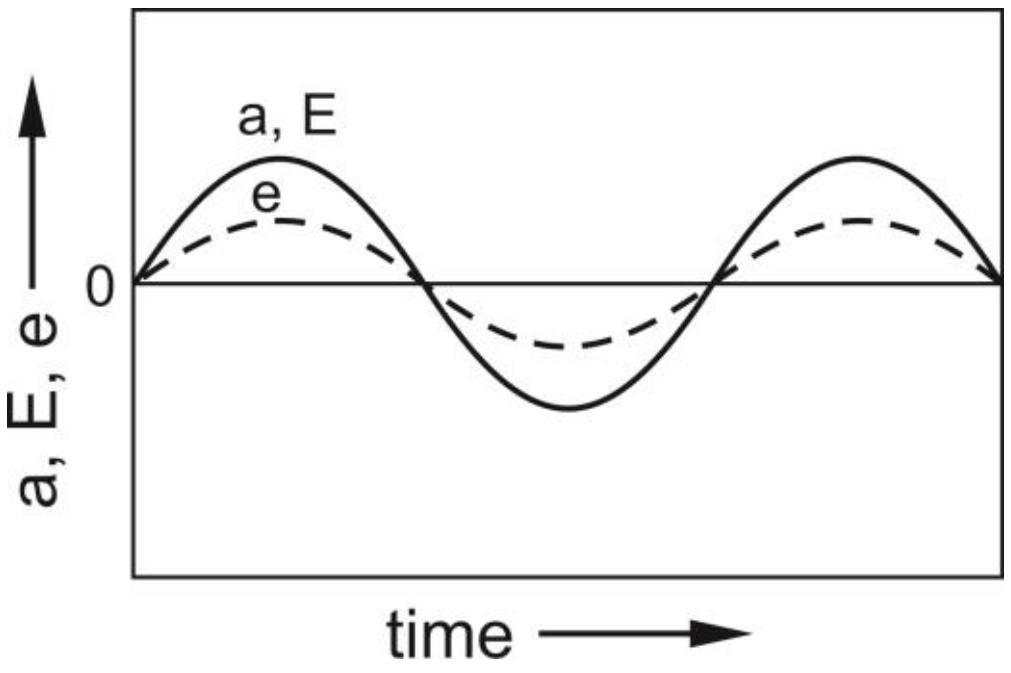

Oscillation of the sociological activity in phase with expected and real success ("seeming and being") as a consequence of the ingredients described in the text. The quantities a, E, e are deviations from the equilibrium state (a = E = e = 0). The "equilibrium" state is characterized by identity of expected and real success ("seeming equals being").

**Figure 3**

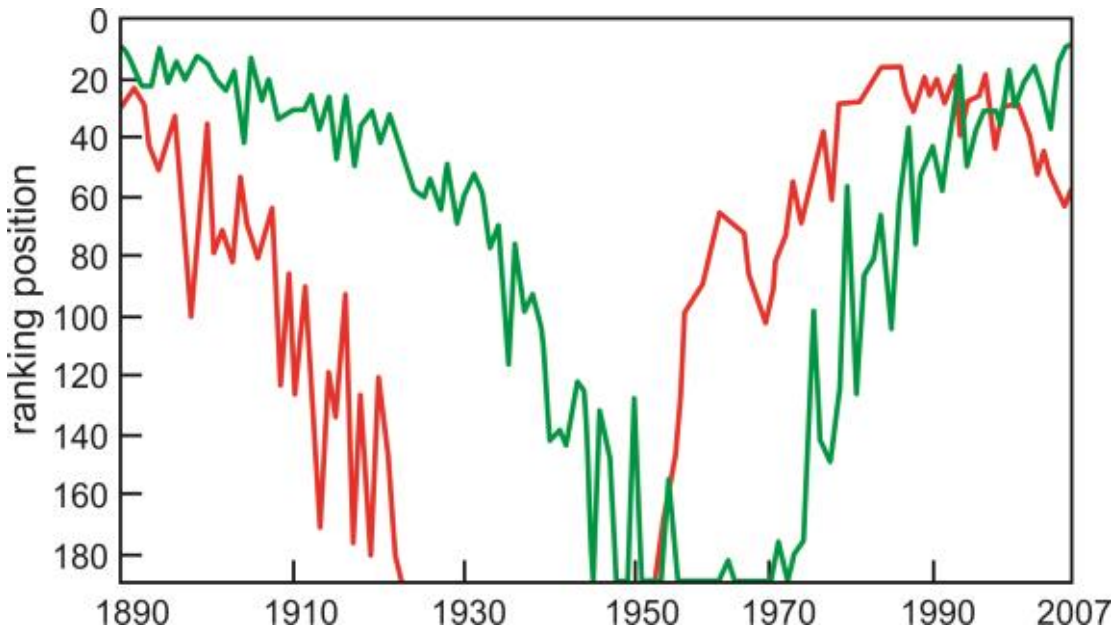

Frequency of the given names “Max” (green) and “Carolin(e)” (or “Karoline”) (red) in Germany for the last 120 years. Decline and revival of the names can be directly explained by the exclusivity effect. (See also suppl. 1.) *Reproduced with permission of knud@beliebte-vornamen.de.*

**Figure 4**

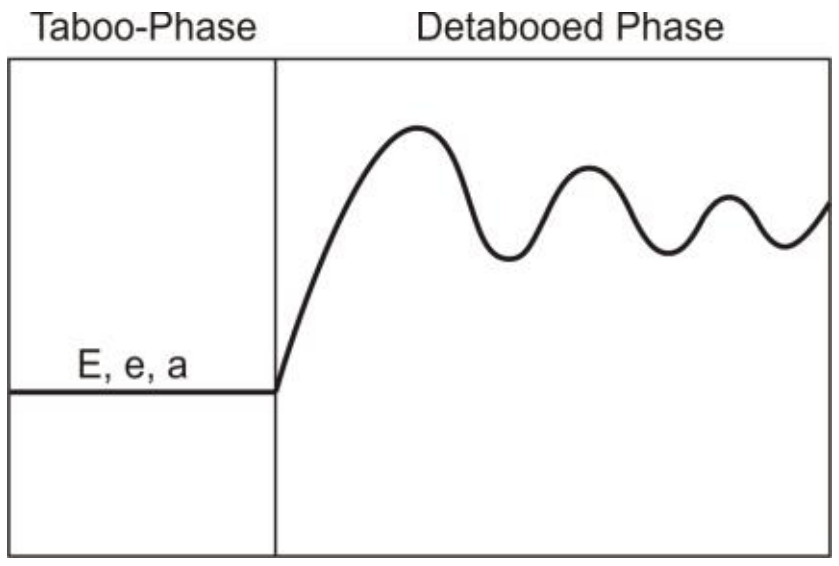

Technology is politically incorrect in the taboo phase. Even though the expectation value of success may be high, there is no reaction in terms of $a$. In the de-tabooed phase, α increases from its (near) zero value to its equilibrium value $\alpha_\infty$. The oscillatory solutions are damped and derived in supplementary 4.

**Figure 5**

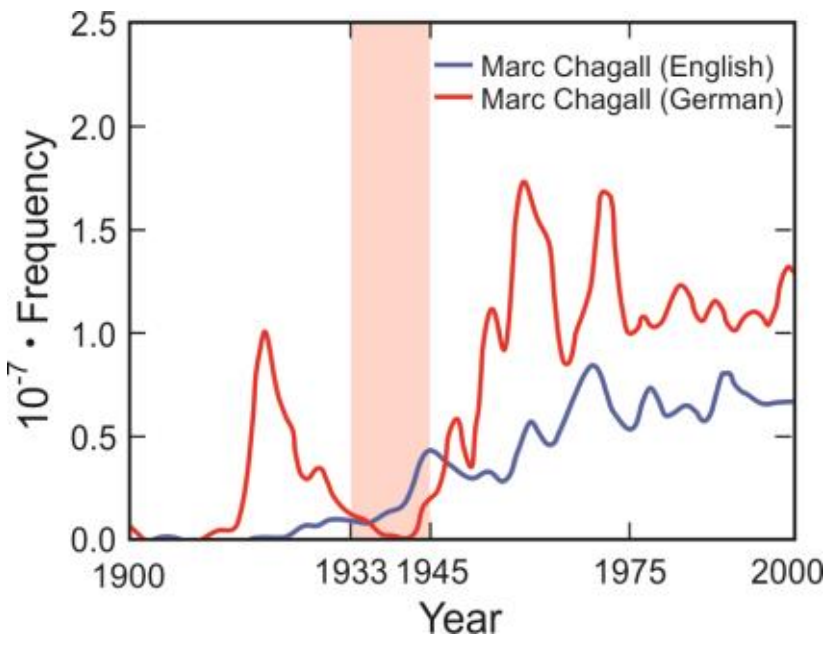

Citations of Marc Chagall in UK and Germany in the last century as a function of time measured by Google-culturomics. *Reproduced with permission of The American Association for the Advancement of Science* (cf. also ref. 23).

**Figure 6**

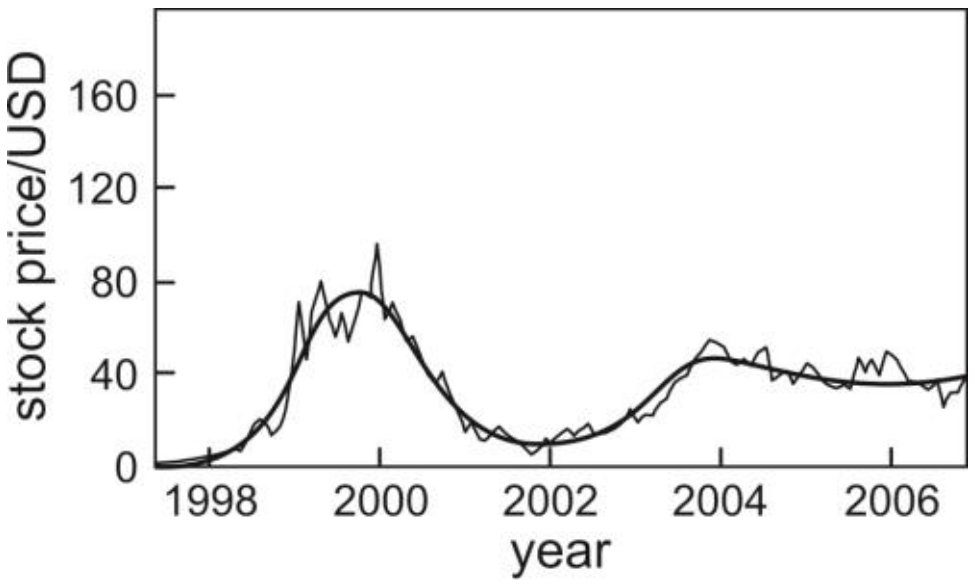

The figure shows the chart plot of Amazon stock price from 1998 to 2007. (In 2007 another upward trend sets in). Realistic activity curves are rather characterized by damped cycles which can be described by superposition of indirect and direct responses or by learning effects.

*Reproduced with permission of Yahoo! Inc. ©2011 Yahoo! Inc. YAHOO! and the YAHOO! logo are registered trademarks of Yahoo! Inc.*

## References

1. Ebeling, W. *Strukturbildung bei irreversiblen Prozessen* (Teubner-Verlag, 1976).

2.. Le Bon, G. *The Crowd: A Study of the Popular Mind* (T. Fisher Unwin, 1896).

3. Noelle-Neumann, E. *Die Schweigespirale. Öffentliche Meinung – unsere soziale Haut* (Langen Müller, 1980).

4. Coleman, J. S. *Introduction to Mathematical Sociology* (The Free Press of Glencoe, Collier-Macmillan Lim., 1904).

5. It is known that in larger groups individual opinions are often subordinated to an overall behavior or even phased.

6. Helbing, D., Farkas, I & Vicsek, T. Simulating dynamical features of escape panic. *Nature* **407**, 487–490 (2000).

7. Radosavljevic, V., Radunovic, D. & Belojevic, G. Epidemics of panic during a bioterrorist attack–A mathematical model. *Medical Hypotheses* **73,** 342–346 (2009).

8. Montroll, E. W. Introduction to Quantitative Aspects of Social Phenomena (Gordon and Breach Science Publ., 1991).

9. Volterra, V. *Opere matematiche* (Accademia Naz. dei Lincei, 1962).

10. Kotomin, E. & Kuzovkov, V.*Modern Aspects of Diffusion Controlled Reactions* (Elsevier, 1996).

11. Weidlich, W. & Haug, E. *Concepts and Models of a Quantitative Sociology* (Springer-Verlag, 1983).

12. Maier, J. *Physical Chemistry of Ionic Materials* (Wiley, 2004).

13. May, R. Time-Delay versus Stability in Population Models with Two or Three Trophic Levels. *Ecology* **54,** 315-325 (1973).

14. Forrester, J. W. *Principles of Systems* (Pegasus Communications, 1968).

15. Berg, E. & Kuhlmann, F. *Systemanalyse und Simulation für Agrarwissenschaftler und Biologen* (Ulmer-Verlag, 1993).

16. Fenn, J. & Raskino, M. *Mastering the Hype Cycle* (Harvard Business Press, 2008 (p. 26)).

17. Pearton, S. *Trust us, we are experts! Mater.Today* **10,** 6-6 (2007).

18. Gomatam, J. & Macdonald, N. Time delays and stability of two competing species. *Math. Biosc.* **24,** 247–256 (1975).

19. Arditi, R. Abillon, J. M., & da Silva, J. V. The Effect of Time-Delay in a Predator-Prey Model. *Math. Biosc.* **33,** 107–120 (1977).

20. Eigen, M. Selforganization of Matter and the Evolution of Biological Marcomolecules. *Naturwiss.* **58,** 465–523 (1971).

21. Landau, L. D. & Lifshitz, E.M. *Course of Theoretical Physics* (Vol **I**, Pergamon Press, 1965).

22. Haken, H. *Synergetics* (Springer-Verlag, 2004).

23. J. B. Michel *et al*., Quantitative Analysis of Culture Using Millions of Digitized Books. *Science* **331,** 176–182 (2011).

**Supplementary Information** is available in the online version of the paper.

**Acknowledgements**

The author thanks K. Funke and in particular E. Kotomin for discussions. He is grateful also to D. Gryaznov and U. Traub for their assistance as far as maple and mathematica is concerned.

## Supplementary 1: The fashion problem

This brief consideration deals with intrinsic formation of fashion cycles. (Extrinsic effects such as periodic supply of fashion articles to avoid marked saturation by fashion industry are not taken account of. Even though important in praxi, they are rather trivial in the present context.) We assume two colors blue (B) and red (R). There are two effects of interest. One is the "exclusivity effect" consisting in B when meeting B will change to R (and vice versa). The other is the "imitation or infection effect" consisting in B getting infected by meeting R will change to R (and vice versa). We concentrate on the first effect and set all the kinetic constants to unity.

Hence, as for the exclusivity effect

$$B + B \rightarrow R + R$$

and

$$R + R \rightarrow B + B$$

are the important reactions which for simplicity run at the same rates. If we assume simple reaction kinetics both b and r (small characters denote the respective concentrations) will monotonically relax from the initial value to the equilibrium value $\hat{b} = \hat{r} = 1/2$.

If we now introduce a memory such that B changes to R if only B has encountered B n-times and vice versa for R; then we need to distinguish between B's that have had such an encounter for the first, second, … n'th time. If, e.g., n=2 we need to consider the events

$$B_0 + B_0 \to B_1 + B_1$$
$$B_0 + B_1 \to B_1 + B_2$$
$$B_0 + B_2 \to B_1 + R_0$$
$$B_1 + B_1 \to B_2 + B_2$$
$$B_1 + B_2 \to B_2 + R_0$$
$$B_2 + B_2 \to R_0 + R_0$$

and the same for R.

Note that these equations do not satisfy microscopic reversibility and should not be used to describe chemical reactions. (In principle the rate constants should depend on the index of the reaction members♦???. This variation is neglected here.) Figure S 1 gives solutions for various n showing that the introduction of a memory leads to damped oscillations (initially 1000 blue and 9000 red particles).

**Fig. S 1**

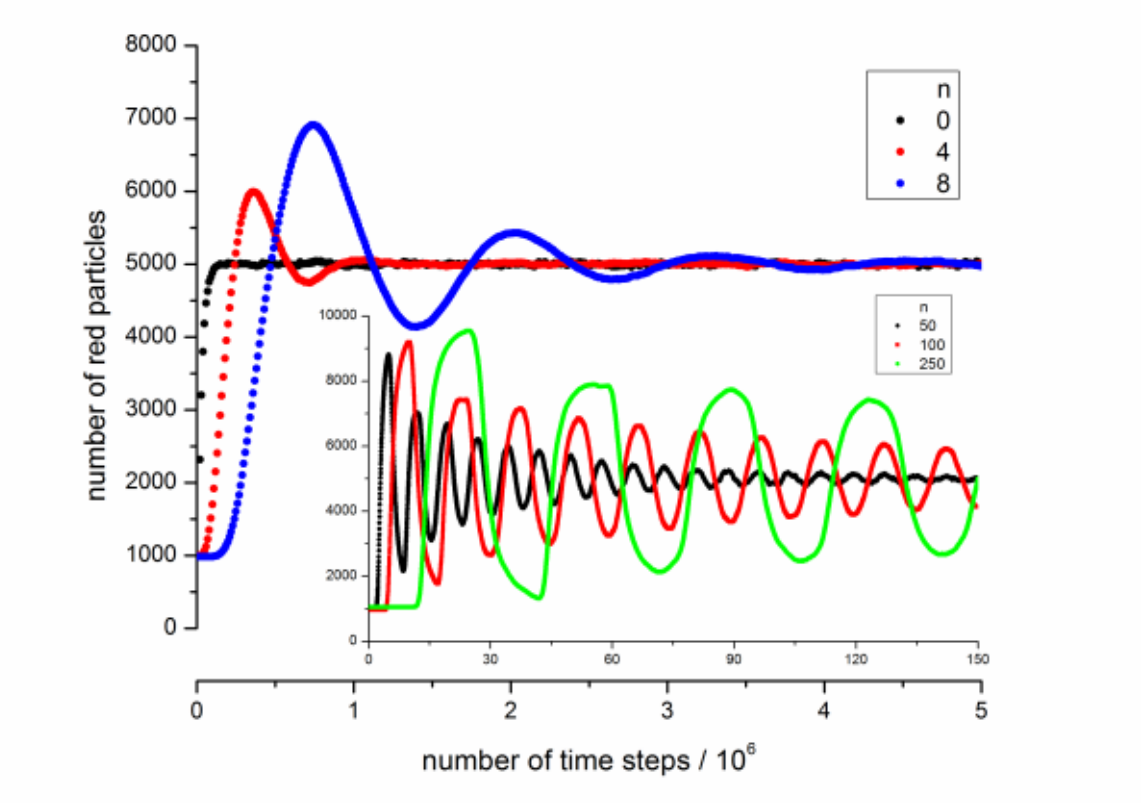

Certainly, the "infection effect", viz. the tendency of B to change to R on meeting R could be easily implemented as well.

The introduction of different species corresponds to a discrete transition from spontaneous to accumulative behavior. Hence damped oscillations occur the behavior of which depends on the strength of the memory and reaction readiness ( $A_1, ..., A_n$ ; $n$ is related with memory and $\alpha$ ).

A simplified version of the above is obtained if we demand B (or R) to react when meeting a "catalyst" K the concentration of which we can keep constant (see Figure 1, main text).

Then, e.g., for n = 2

$$\begin{aligned} B_0 + K &\rightarrow B_1 + K \\ B_1 + K &\rightarrow B_2 + K \\ B_2 + K &\rightarrow R_0 + K \end{aligned}$$

and the same for R.

Even though these equations are now (since k = const) of first order, the general picture is not very different.

The reaction without memory

$$B + K \rightleftarrows R + K$$

is describable by an exponential function that satisfies the rate law

$$\text{rate} \propto b\text{-}\widehat{b} = -\left(r - \widehat{r}\right)$$

As described in the main text, in order to treat more general situations, we proceed to a description that is not necessarily based on collision events but includes memory by demanding

$$\text{rate} \propto \int \left(b\text{-}\widehat{b}\right) dt.$$

**Supplementary 2:**

*Solutions for* $\ddot{a}\left(\tau\right) = a^n\left(\tau\right) - a^m\left(\tau\right)$ *with selected n, m*:

a) The differential equation $\ddot{a}\left(\tau\right) = a^{1/2}\left(\tau\right) - a\left(\tau\right)$ has the solution

$$\tau = -2\,ArcTan\left[\frac{\left(-2+3\sqrt{a}\right)\sqrt{4a^{3/2}-3a^2}}{\sqrt{3\left(-4+3\sqrt{a}\right)a}}\right]$$

or explicitly for trivial boundary conditions

$$a(\tau) = a_1(\tau) @ a_2(\tau_2)$$

$$a_1 = \frac{4\left(1+3Tan\left[\frac{t}{2}\right]^2+2Tan\left[\frac{t}{2}\right]^4-\ 2\sqrt{Tan\left[\frac{t}{2}\right]^2+3Tan\left[\frac{t}{2}\right]^4+3Tan\left[\frac{t}{2}\right]^6+Tan\left[\frac{t}{2}\right]^8}\right)}{9\left(1+2Tan\left[\frac{t}{2}\right]^2+Tan\left[\frac{t}{2}\right]^4\right)}$$

$$a_2 = \frac{4\left(1+3Tan\left[\frac{t}{2}\right]^2+2Tan\left[\frac{t}{2}\right]^4+\ 2\sqrt{Tan\left[\frac{t}{2}\right]^2+3Tan\left[\frac{t}{2}\right]^4+3Tan\left[\frac{t}{2}\right]^6+Tan\left[\frac{t}{2}\right]^8}\right)}{9\left(1+2Tan\left[\frac{t}{2}\right]^2+Tan\left[\frac{t}{2}\right]^4\right)}$$

as sketched in Fig. supplementary 2A.

**Fig. S 2A**

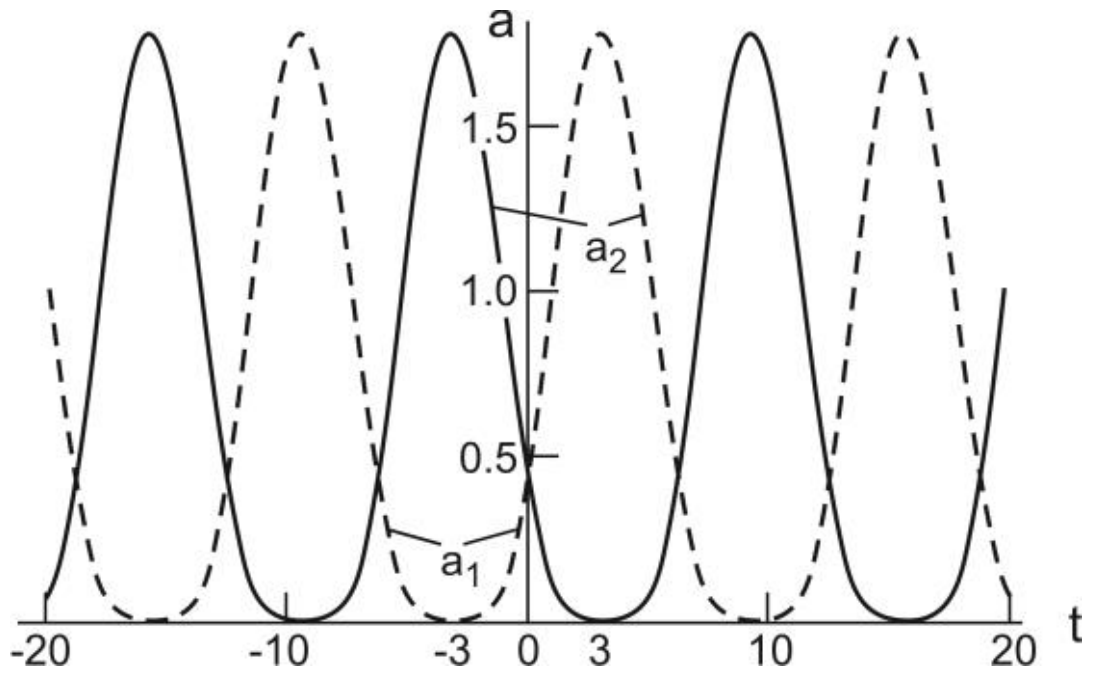

The intercepts with the $\tau$ -axis are odd multiples of π; the intercept with the *a*-axis is 4/9 . The inflection point is at $a$ = 1; the maximum is 16/9. The oscillating functions are composites of $a_1$ and $a_2$.

b) The solution for $\ddot{a}(\tau) = a^2(\tau) - a(\tau)$ is for trivial boundary conditions

$$a(\tau) = \frac{3}{2}\left(1 + \left(Tan\left(\frac{\tau}{2}\right)\right)^2\right)$$

The graph is sketched in Fig. suppl. 2B.

**Fig. S 2B**

$$plot\left(\frac{3}{2}\cdot\left(1 + \left(\tan\left(\frac{\text{tau}}{2}\right)\right)^2\right), \text{tau} = -15\,..15, -10\,..50\right);$$

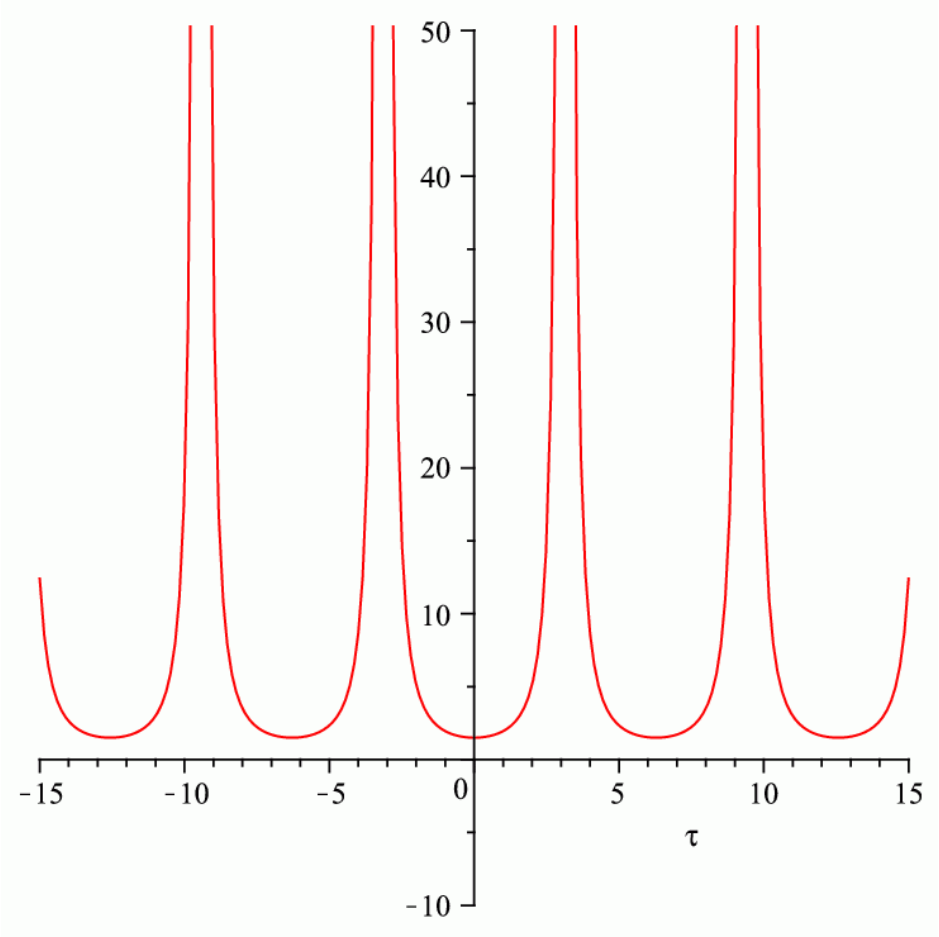

c) The differential equation

$$\ddot{a}(\tau) = a^{3/2}(\tau) - a(\tau)$$

delivers a periodical but non oscillating function

$$a(\tau) = \frac{25}{16}\left[1 + \tan\left[\frac{\tau}{4}\right]^2 + \tan\left[\frac{\tau}{4}\right]^4\right]$$

The $a(\tau)$ curves are given in Fig. supplementary 2 C.

**Fig. S 2C**

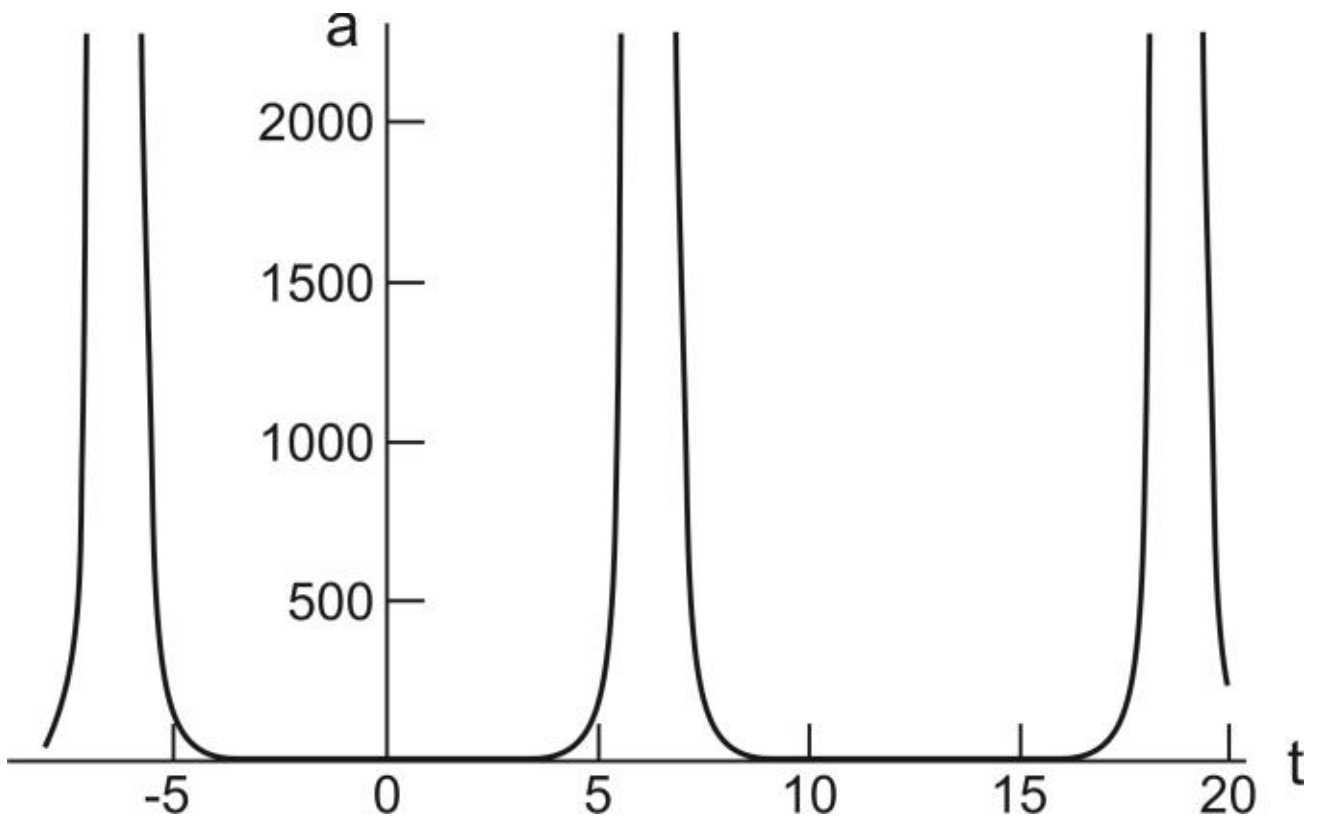

d) It is interesting to compare the solutions for $\ddot{a}(\tau) = a^n(\tau) - a^m(\tau)$ with the spontaneous reaction

$\dot{a}(\tau) = a^n(\tau) - a^m(\tau)$. Let us first consider

$\dot{a}(\tau) = a(\tau) - a^2(\tau)$

The solution for $a(0) \neq 1$ is

$$a(\tau) = \frac{\frac{a(0)}{1-a(0)}\exp\tau}{1+\frac{a(0)}{1-a(0)}\exp\tau}$$

For $a < 1$ the first term predominates and $a$ increases. For $a > 1$ the second term predominates and $a$ decreases. Yet, unlike the accumulative response, oscillations do not occur, rather the value $a = 1$ forms a boundary for the solutions. If $a(\tau = 0) \equiv \mathrm{a}(0) > 1$, the $a$-function stays always greater than unity, while it stays below unity if $a(0) < 1$. See Fig. suppl. 2D. A similar compartmentation occurs for the solution of

$\dot{a}(\tau) = a^2(\tau) - a(\tau)$

which reads for $a(0) \neq 1$

$$a(\tau) = -\frac{\frac{a(0)}{a(0)-1}}{\exp\tau - \frac{a(0)}{a(0)-1}}$$

**Fig. S 2D**

$$plot\left(\left\{\frac{4}{3}\cdot\frac{\exp(\text{tau})}{1+\left(\frac{4}{3}\right)\cdot\exp(\text{tau})}, -\frac{\exp(\text{tau})}{1-\exp(\text{tau})}\right\}, \text{tau}=-5..5, -5..5\right);$$

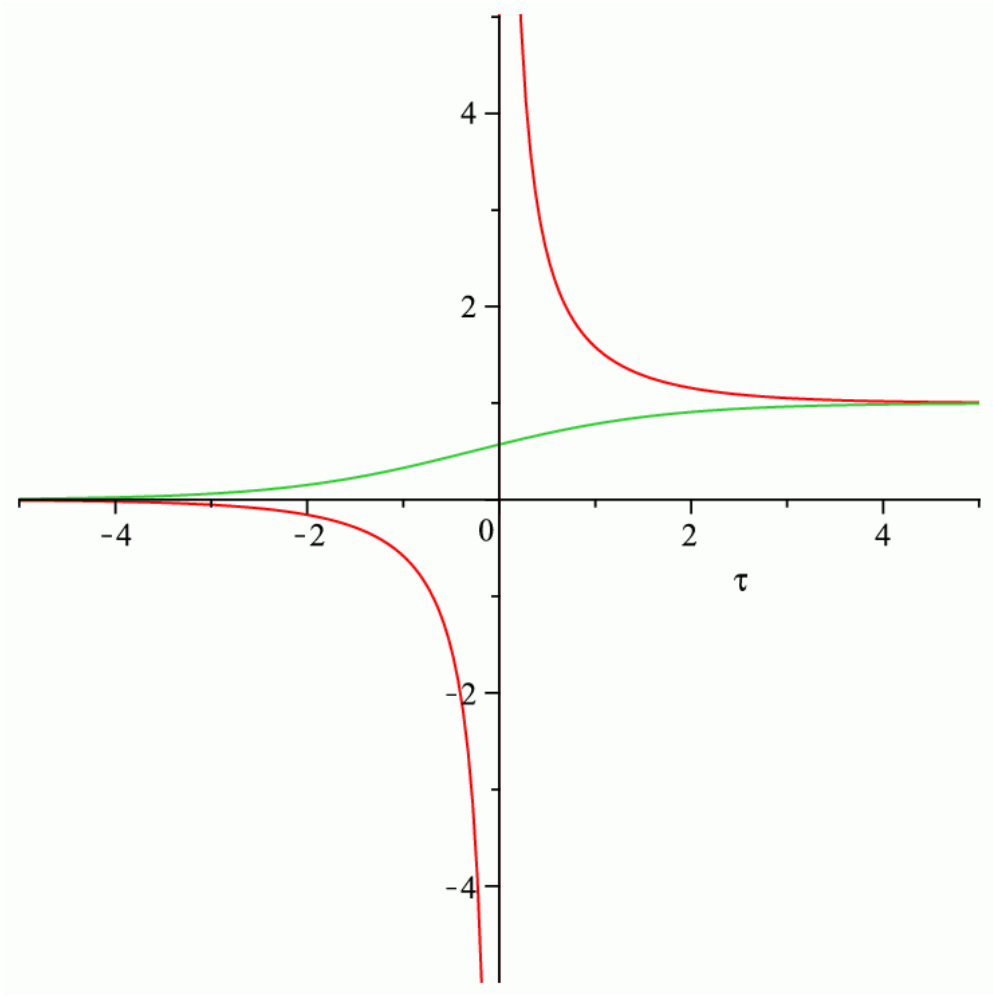

and is displayed in Fig. suppl. 2E, which may be compared with Fig. 2B.

Such a compartmentation is dissolved for an accumulative response.

**Fig. S 2E**

$$plot\left(\left\{-\frac{4}{3}\cdot\left(\exp(tau)-\frac{4}{3}\right)^{-1}, (\exp(tau)+1)^{-1}\right\}, tau=-5..5, -5..5\right);$$

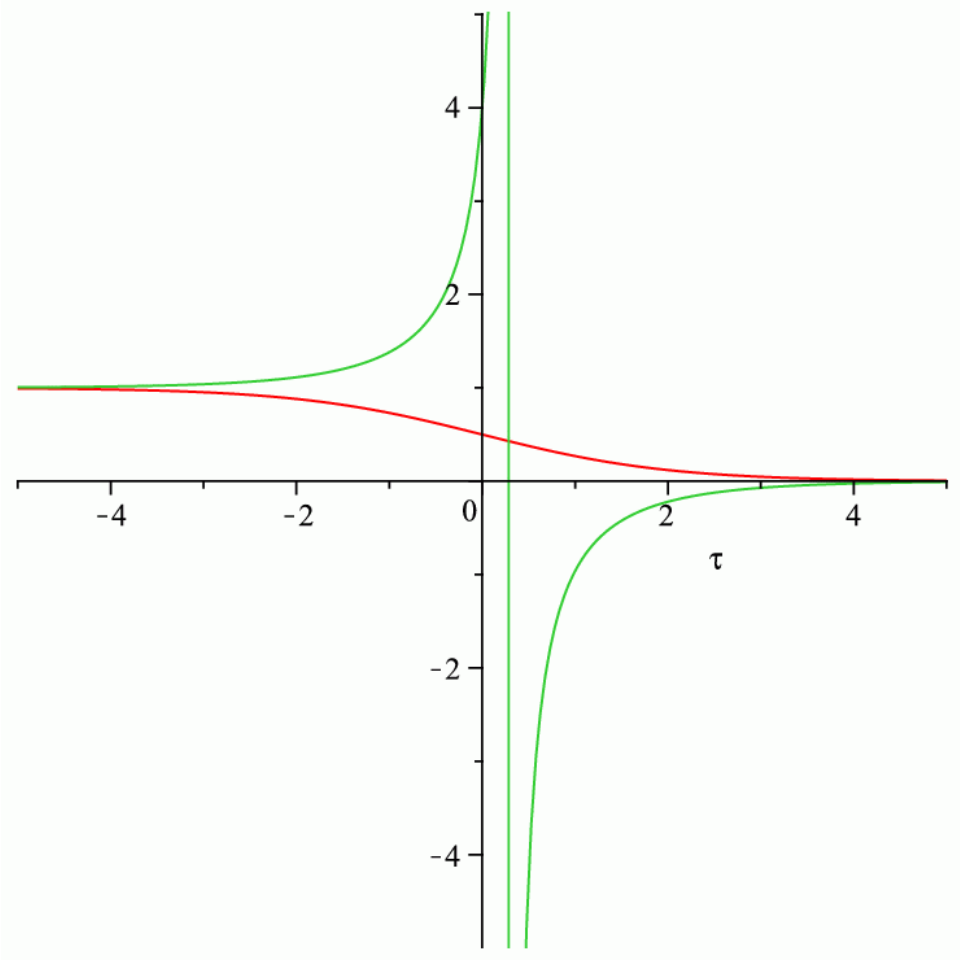

## Supplementary 3: Taboo-phase and breakdown

Let us for simplicity assume direct, spontaneous reaction (rather than an accumulative one) what enables us to adopt the simple formalism of chemical kinetics.

The taboo-phase (or phase of directorship) is determined by the fact that any unwanted action (activity) A must rapidly decay (large reaction constant $\vec{k}_D$) according to

$$\mathrm{A} \xrightarrow{\vec{k}_D} \mathrm{Nil}. \tag{4}$$

Nonetheless there is also a growth mechanism that is enabled wherever a precursor $\mathrm{A}^*$ is met ($\mathrm{A}^*$ is an activity in a critical pre-stage)

$$\mathrm{A}+\mathrm{A}^* \xrightarrow{\vec{k}_G} 2\mathrm{A}. \tag{5}$$

The overall rate is $[\mathrm{A}]\left(\vec{k}_G\left[\mathrm{A}^*\right]-\vec{k}_D\right)$. At a high enough number density of $\mathrm{A}^*$ $\left(\text{i.e. of } \left[\mathrm{A}^*\right]\right)$, namely for $\vec{k}_G\left[\mathrm{A}^*\right]>\vec{k}_D$, channel 2 dominates: A grows finally leading to a breakdown of the taboo-phase. The fate of it is determined by the ratio $\frac{\vec{k}_G\left[\mathrm{A}^*\right]}{\vec{k}_D}$. The situation becomes very interesting, if several activities compete with each other and compete about an exhaustible number of $\mathrm{A}^*$. This leads automatically to a selection problem in which the growth (survival) depends on other activities. Incorporation of memory effects to this treatment which follows ref. 20 would lead to analogous solutions for sociological problems.

**Supplementary 4: Solutions including pre-phases**

**Supplementary 4a**

$$dgl := diff(y(x), x, x) + 100\cdot\left(1 - \exp\left(-\frac{x}{10}\right)\right)\cdot y(x) = 0;$$

$$\frac{\mathrm{d}^2}{\mathrm{d}x^2}\, y(x) + 100\left(1 - \mathrm{e}^{-\frac{1}{10}x}\right) y(x) = 0$$

$$dsolve(\{dgl, y(0) = -1000, \mathrm{D}(y)(0) = 0\}, y(x));$$

$$y(x) = \Big(1000\,(-\mathrm{BesselK}(1 + 200\,I, 200) + I\,\mathrm{BesselK}(200\,I, 200))\,\mathrm{BesselI}\Big(200\,I, 200\,e^{-\frac{1}{20}x}\Big)\Big) \Big/ (\mathrm{BesselK}(200\,I, 200)\,\mathrm{BesselI}(1 + 200\,I, 200) + \mathrm{BesselK}(1 + 200\,I, 200)\,\mathrm{BesselI}(200\,I, 200)) - \Big(1000\,(\mathrm{BesselI}(1 + 200\,I, 200) + I\,\mathrm{BesselI}(200\,I, 200))\,\mathrm{BesselK}\Big(200\,I, 200\,e^{-\frac{1}{20}x}\Big)\Big) \Big/ (\mathrm{BesselK}(200\,I, 200)\,\mathrm{BesselI}(1 + 200\,I, 200) + \mathrm{BesselK}(1 + 200\,I, 200)\,\mathrm{BesselI}(200\,I, 200))$$

*funcy* := *unapply*(*rhs*(*dsolve*({*dgl*, *y*(0) = −1000, D(*y*)(0) = 0}, *y*(*x*))), *x*);

$$x \rightarrow \Big(1000\,(-\mathrm{BesselK}(1 + 200\,I, 200) + I\,\mathrm{BesselK}(200\,I, 200))\,\mathrm{BesselI}\Big(200\,I, 200\,e^{-\frac{1}{20}x}\Big)\Big) \Big/ (\mathrm{BesselK}(200\,I, 200)\,\mathrm{BesselI}(1 + 200\,I, 200) + \mathrm{BesselK}(1 + 200\,I, 200)\,\mathrm{BesselI}(200\,I, 200)) - \Big(1000\,(\mathrm{BesselI}(1 + 200\,I, 200) + I\,\mathrm{BesselI}(200\,I, 200))\,\mathrm{BesselK}\Big(200\,I, 200\,e^{-\frac{1}{20}x}\Big)\Big) \Big/ (\mathrm{BesselK}(200\,I, 200)\,\mathrm{BesselI}(1 + 200\,I, 200) + \mathrm{BesselK}(1 + 200\,I, 200)\,\mathrm{BesselI}(200\,I, 200))$$

**Fig. S 4a**

*plot*(*funcy*(*x*), *x* = −0.5 ..5);

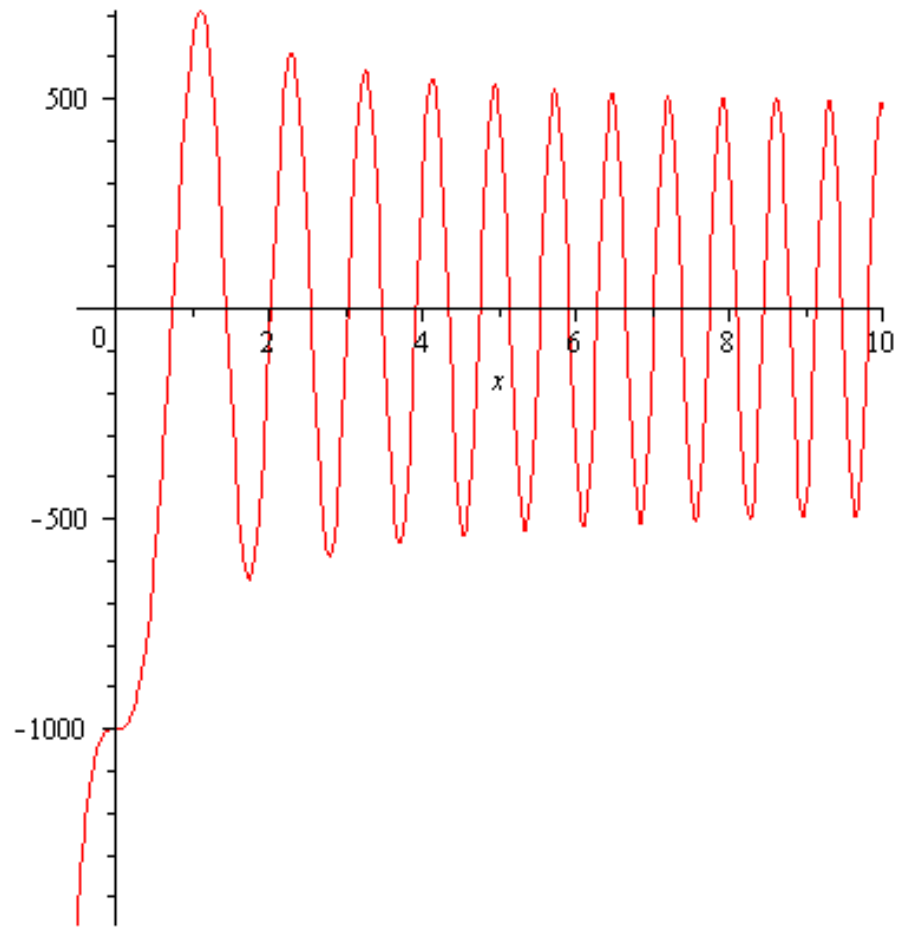

**Supplementary 4b**

$dgl := diff(y(x), x, x) + \left(2 - 1.99 \cdot \exp\left(-\frac{x}{5}\right)\right)^{-1} \cdot y(x) = 0;$

$$\frac{d^2}{dx^2} y(x) + \frac{y(x)}{2 - 1.99 e^{-\frac{1}{5} x}} = 0 \qquad \textbf{(1)}$$

$dsolve(\{dgl, y(0) = 10, D(y)(0) = 0\}, y(x));$

$$y(x) = -\Bigg(10 \bigg(102000\, I\, \mathrm{hypergeom}\left(\left[1 + \frac{5}{2} I\sqrt{2}, 1 + \frac{5}{2} I\sqrt{2}\right], \left[1 + 5 I\sqrt{2}\right], \frac{199}{200}\right)\sqrt{2} - 8119200\, \mathrm{hypergeom}\left(\left[1 + \frac{5}{2} I\sqrt{2}, 1 + \frac{5}{2} I\sqrt{2}\right], \left[1 + 5 I\sqrt{2}\right], \frac{199}{200}\right) + 30646\, \mathrm{hypergeom}\left(\left[2 + \frac{5}{2} I\sqrt{2}, 2 + \frac{5}{2} I\sqrt{2}\right], \left[2 + 5 I\sqrt{2}\right], \frac{199}{200}\right) + 49750\, I\, \mathrm{hypergeom}\left(\left[2 + \frac{5}{2} I\sqrt{2}, 2 + \frac{5}{2} I\sqrt{2}\right], \left[2 + 5 I\sqrt{2}\right], \frac{199}{200}\right)\sqrt{2}\bigg) \left(-200 e^{\frac{1}{2} I\sqrt{2} x} + 199 e^{\frac{1}{2} I\sqrt{2} x - \frac{1}{5} x}\right) \mathrm{hypergeom}\left(\left[1 - \frac{5}{2} I\sqrt{2}, 1 - \frac{5}{2} I\sqrt{2}\right], \left[1 - 5 I\sqrt{2}\right], \frac{199}{200} e^{-\frac{1}{5} x}\right)\Bigg) \Bigg/ \Bigg(204000\, I\, \mathrm{hypergeom}\left(\left[1 + \frac{5}{2} I\sqrt{2}, 1 + \frac{5}{2} I\sqrt{2}\right], \left[1 + 5 I\sqrt{2}\right], \frac{199}{200}\right)\sqrt{2}\, \mathrm{hypergeom}\left(\left[1 - \frac{5}{2} I\sqrt{2}, 1 - \frac{5}{2} I\sqrt{2}\right], \left[1 - 5 I\sqrt{2}\right], \frac{199}{200}\right) - 30646\, \mathrm{hypergeom}\left(\left[1 + \frac{5}{2} I\sqrt{2}, 1 + \frac{5}{2} I\sqrt{2}\right], \left[1 + 5 I\sqrt{2}\right], \frac{199}{200}\right) \mathrm{hypergeom}\left(\left[2 - \frac{5}{2} I\sqrt{2}, 2 - \frac{5}{2} I\sqrt{2}\right], \left[2 - 5 I\sqrt{2}\right], \frac{199}{200}\right) + 49750\, I\, \mathrm{hypergeom}\left(\left[1 + \frac{5}{2} I\sqrt{2}, 1\right.\right. \qquad \textbf{(2)}$$

$$+\frac{5}{2}\,I\sqrt{2}],\ [1+5\,I\sqrt{2}],\ \frac{199}{200})\ \mathrm{hypergeom}([2-\frac{5}{2}\,I\sqrt{2},\ 2-\frac{5}{2}\,I\sqrt{2}],\ [2-5\,I\sqrt{2}],\ \frac{199}{200})\sqrt{2}$$

$$+30646\,\mathrm{hypergeom}([2+\frac{5}{2}\,I\sqrt{2},\ 2+\frac{5}{2}\,I\sqrt{2}],\ [2+5\,I\sqrt{2}],\ \frac{199}{200})\ \mathrm{hypergeom}([1-\frac{5}{2}\,I\sqrt{2},\ 1-\frac{5}{2}\,I\sqrt{2}],\ [1$$

$$-5\,I\sqrt{2}],\ \frac{199}{200})+49750\,I\,\mathrm{hypergeom}([2+\frac{5}{2}\,I\sqrt{2},\ 2+\frac{5}{2}\,I\sqrt{2}],\ [2+5\,I\sqrt{2}],\ \frac{199}{200})\sqrt{2}\ \mathrm{hypergeom}([1$$

$$-\frac{5}{2}\,I\sqrt{2},\ 1-\frac{5}{2}\,I\sqrt{2}],\ [1-5\,I\sqrt{2}],\ \frac{199}{200}))-(10\,(102000\,I\,\mathrm{hypergeom}([1-\frac{5}{2}\,I\sqrt{2},\ 1-\frac{5}{2}\,I\sqrt{2}],$$

$$[1-5\,I\sqrt{2}],\ \frac{199}{200})\sqrt{2}+8119200\,\mathrm{hypergeom}([1-\frac{5}{2}\,I\sqrt{2},\ 1-\frac{5}{2}\,I\sqrt{2}],\ [1-5\,I\sqrt{2}],\ \frac{199}{200})$$

$$-30646\,\mathrm{hypergeom}([2-\frac{5}{2}\,I\sqrt{2},\ 2-\frac{5}{2}\,I\sqrt{2}],\ [2-5\,I\sqrt{2}],\ \frac{199}{200})+49750\,I\,\mathrm{hypergeom}([2-\frac{5}{2}\,I\sqrt{2},\ 2$$

$$-\frac{5}{2}\,I\sqrt{2}],\ [2-5\,I\sqrt{2}],\ \frac{199}{200})\sqrt{2})\,(-200\,e^{-\frac{1}{2}I\sqrt{2}\,x}+199\,e^{-\frac{1}{2}I\sqrt{2}\,x-\frac{1}{5}x})\ \mathrm{hypergeom}([1+\frac{5}{2}\,I\sqrt{2},\ 1$$

$$+\frac{5}{2}\,I\sqrt{2}],\ [1+5\,I\sqrt{2}],\ \frac{199}{200}\,e^{-\frac{1}{5}x}))\ /\ (204000\,I\,\mathrm{hypergeom}([1+\frac{5}{2}\,I\sqrt{2},\ 1+\frac{5}{2}\,I\sqrt{2}],\ [1+5\,I\sqrt{2}],$$

$$\frac{199}{200})\sqrt{2}\ \mathrm{hypergeom}([1-\frac{5}{2}\,I\sqrt{2},\ 1-\frac{5}{2}\,I\sqrt{2}],\ [1-5\,I\sqrt{2}],\ \frac{199}{200})-30646\,\mathrm{hypergeom}([1+\frac{5}{2}\,I\sqrt{2},\ 1$$

$$+\frac{5}{2}\,I\sqrt{2}],\ [1+5\,I\sqrt{2}],\ \frac{199}{200})\ \mathrm{hypergeom}([2-\frac{5}{2}\,I\sqrt{2},\ 2-\frac{5}{2}\,I\sqrt{2}],\ [2-5\,I\sqrt{2}],\ \frac{199}{200})$$

$$+49750\,I\,\mathrm{hypergeom}([1+\frac{5}{2}\,I\sqrt{2},\ 1+\frac{5}{2}\,I\sqrt{2}],\ [1+5\,I\sqrt{2}],\ \frac{199}{200})\ \mathrm{hypergeom}([2-\frac{5}{2}\,I\sqrt{2},\ 2-\frac{5}{2}\,I\sqrt{2}],$$

$$[2-5\,I\sqrt{2}],\ \frac{199}{200})\sqrt{2}+30646\,\mathrm{hypergeom}([2+\frac{5}{2}\,I\sqrt{2},\ 2+\frac{5}{2}\,I\sqrt{2}],\ [2+5\,I\sqrt{2}],\ \frac{199}{200})\ \mathrm{hypergeom}([1$$

$$-\frac{5}{2}\,I\sqrt{2},\ 1-\frac{5}{2}\,I\sqrt{2}],\ [1-5\,I\sqrt{2}],\ \frac{199}{200})+49750\,I\,\mathrm{hypergeom}([2+\frac{5}{2}\,I\sqrt{2},\ 2+\frac{5}{2}\,I\sqrt{2}],\ [2+5\,I\sqrt{2}],$$

$$\frac{199}{200})\sqrt{2}\ \mathrm{hypergeom}([1-\frac{5}{2}\,I\sqrt{2},\ 1-\frac{5}{2}\,I\sqrt{2}],\ [1-5\,I\sqrt{2}],\ \frac{199}{200}))$$

*funcy* := *unapply*(*rhs*(*dsolve*({*dgl*, *y*(0) = 10, D(*y*)(0) = 0}, *y*(*x*))), *x*);

$$x\rightarrow-(10\,(102000\,I\,\mathrm{hypergeom}([1+\frac{5}{2}\,I\sqrt{2},\ 1+\frac{5}{2}\,I\sqrt{2}],\ [1+5\,I\sqrt{2}],\ \frac{199}{200})\sqrt{2}-8119200\,\mathrm{hypergeom}([1 \qquad \textbf{(3)}$$

$$+\frac{5}{2}\,\mathrm{I}\sqrt{2},\,1+\frac{5}{2}\,\mathrm{I}\sqrt{2}\Big],\,\Big[1+5\,\mathrm{I}\sqrt{2}\Big],\,\frac{199}{200}\Big)+30646\,\mathrm{hypergeom}\Big(\Big[2+\frac{5}{2}\,\mathrm{I}\sqrt{2},\,2+\frac{5}{2}\,\mathrm{I}\sqrt{2}\Big],\,\Big[2+5\,\mathrm{I}\sqrt{2}\Big],$$

$$\frac{199}{200}\Big)+49750\,\mathrm{I}\,\mathrm{hypergeom}\Big(\Big[2+\frac{5}{2}\,\mathrm{I}\sqrt{2},\,2+\frac{5}{2}\,\mathrm{I}\sqrt{2}\Big],\,\Big[2+5\,\mathrm{I}\sqrt{2}\Big],\,\frac{199}{200}\Big)\sqrt{2}\Big)\Big(-200\,e^{\frac{1}{2}\,\mathrm{I}\sqrt{2}\,x}$$

$$+199\,e^{\frac{1}{2}\,\mathrm{I}\sqrt{2}\,x-\frac{1}{5}\,x}\Big)\,\mathrm{hypergeom}\Big(\Big[1-\frac{5}{2}\,\mathrm{I}\sqrt{2},\,1-\frac{5}{2}\,\mathrm{I}\sqrt{2}\Big],\,\Big[1-5\,\mathrm{I}\sqrt{2}\Big],\,\frac{199}{200}\,e^{-\frac{1}{5}\,x}\Big)\Big)\Big)\Big/$$

$$\Big(204000\,\mathrm{I}\,\mathrm{hypergeom}\Big(\Big[1+\frac{5}{2}\,\mathrm{I}\sqrt{2},\,1+\frac{5}{2}\,\mathrm{I}\sqrt{2}\Big],\,\Big[1+5\,\mathrm{I}\sqrt{2}\Big],\,\frac{199}{200}\Big)\sqrt{2}\,\mathrm{hypergeom}\Big(\Big[1-\frac{5}{2}\,\mathrm{I}\sqrt{2},\,1$$

$$-\frac{5}{2}\,\mathrm{I}\sqrt{2}\Big],\,\Big[1-5\,\mathrm{I}\sqrt{2}\Big],\,\frac{199}{200}\Big)-30646\,\mathrm{hypergeom}\Big(\Big[1+\frac{5}{2}\,\mathrm{I}\sqrt{2},\,1+\frac{5}{2}\,\mathrm{I}\sqrt{2}\Big],\,\Big[1+5\,\mathrm{I}\sqrt{2}\Big],$$

$$\frac{199}{200}\Big)\,\mathrm{hypergeom}\Big(\Big[2-\frac{5}{2}\,\mathrm{I}\sqrt{2},\,2-\frac{5}{2}\,\mathrm{I}\sqrt{2}\Big],\,\Big[2-5\,\mathrm{I}\sqrt{2}\Big],\,\frac{199}{200}\Big)+49750\,\mathrm{I}\,\mathrm{hypergeom}\Big(\Big[1+\frac{5}{2}\,\mathrm{I}\sqrt{2},\,1$$

$$+\frac{5}{2}\,\mathrm{I}\sqrt{2}\Big],\,\Big[1+5\,\mathrm{I}\sqrt{2}\Big],\,\frac{199}{200}\Big)\,\mathrm{hypergeom}\Big(\Big[2-\frac{5}{2}\,\mathrm{I}\sqrt{2},\,2-\frac{5}{2}\,\mathrm{I}\sqrt{2}\Big],\,\Big[2-5\,\mathrm{I}\sqrt{2}\Big],\,\frac{199}{200}\Big)\sqrt{2}$$

$$+30646\,\mathrm{hypergeom}\Big(\Big[2+\frac{5}{2}\,\mathrm{I}\sqrt{2},\,2+\frac{5}{2}\,\mathrm{I}\sqrt{2}\Big],\,\Big[2+5\,\mathrm{I}\sqrt{2}\Big],\,\frac{199}{200}\Big)\,\mathrm{hypergeom}\Big(\Big[1-\frac{5}{2}\,\mathrm{I}\sqrt{2},\,1-\frac{5}{2}\,\mathrm{I}\sqrt{2}\Big],\,\Big[1$$

$$-5\,\mathrm{I}\sqrt{2}\Big],\,\frac{199}{200}\Big)+49750\,\mathrm{I}\,\mathrm{hypergeom}\Big(\Big[2+\frac{5}{2}\,\mathrm{I}\sqrt{2},\,2+\frac{5}{2}\,\mathrm{I}\sqrt{2}\Big],\,\Big[2+5\,\mathrm{I}\sqrt{2}\Big],\,\frac{199}{200}\Big)\sqrt{2}\,\mathrm{hypergeom}\Big(\Big[1$$

$$-\frac{5}{2}\,\mathrm{I}\sqrt{2},\,1-\frac{5}{2}\,\mathrm{I}\sqrt{2}\Big],\,\Big[1-5\,\mathrm{I}\sqrt{2}\Big],\,\frac{199}{200}\Big)\Big)-\Big(10\,\Big(102000\,\mathrm{I}\,\mathrm{hypergeom}\Big(\Big[1-\frac{5}{2}\,\mathrm{I}\sqrt{2},\,1-\frac{5}{2}\,\mathrm{I}\sqrt{2}\Big],$$

$$\Big[1-5\,\mathrm{I}\sqrt{2}\Big],\,\frac{199}{200}\Big)\sqrt{2}+8119200\,\mathrm{hypergeom}\Big(\Big[1-\frac{5}{2}\,\mathrm{I}\sqrt{2},\,1-\frac{5}{2}\,\mathrm{I}\sqrt{2}\Big],\,\Big[1-5\,\mathrm{I}\sqrt{2}\Big],\,\frac{199}{200}\Big)$$

$$-30646\,\mathrm{hypergeom}\Big(\Big[2-\frac{5}{2}\,\mathrm{I}\sqrt{2},\,2-\frac{5}{2}\,\mathrm{I}\sqrt{2}\Big],\,\Big[2-5\,\mathrm{I}\sqrt{2}\Big],\,\frac{199}{200}\Big)+49750\,\mathrm{I}\,\mathrm{hypergeom}\Big(\Big[2-\frac{5}{2}\,\mathrm{I}\sqrt{2},\,2$$

$$-\frac{5}{2}\,\mathrm{I}\sqrt{2}\Big],\,\Big[2-5\,\mathrm{I}\sqrt{2}\Big],\,\frac{199}{200}\Big)\sqrt{2}\Big)\Big(-200\,e^{-\frac{1}{2}\,\mathrm{I}\sqrt{2}\,x}+199\,e^{-\frac{1}{2}\,\mathrm{I}\sqrt{2}\,x-\frac{1}{5}\,x}\Big)\,\mathrm{hypergeom}\Big(\Big[1+\frac{5}{2}\,\mathrm{I}\sqrt{2},\,1$$

$$+\frac{5}{2}\,\mathrm{I}\sqrt{2}\Big],\,\Big[1+5\,\mathrm{I}\sqrt{2}\Big],\,\frac{199}{200}\,e^{-\frac{1}{5}\,x}\Big)\Big)\Big/\Big(204000\,\mathrm{I}\,\mathrm{hypergeom}\Big(\Big[1+\frac{5}{2}\,\mathrm{I}\sqrt{2},\,1+\frac{5}{2}\,\mathrm{I}\sqrt{2}\Big],\,\Big[1+5\,\mathrm{I}\sqrt{2}\Big],$$

$$\frac{199}{200}\Big)\sqrt{2}\,\mathrm{hypergeom}\Big(\Big[1-\frac{5}{2}\,\mathrm{I}\sqrt{2},\,1-\frac{5}{2}\,\mathrm{I}\sqrt{2}\Big],\,\Big[1-5\,\mathrm{I}\sqrt{2}\Big],\,\frac{199}{200}\Big)-30646\,\mathrm{hypergeom}\Big(\Big[1+\frac{5}{2}\,\mathrm{I}\sqrt{2},\,1$$

$$+\frac{5}{2}\,\mathrm{I}\sqrt{2}\Big],\,\Big[1+5\,\mathrm{I}\sqrt{2}\Big],\,\frac{199}{200}\Big)\,\mathrm{hypergeom}\Big(\Big[2-\frac{5}{2}\,\mathrm{I}\sqrt{2},\,2-\frac{5}{2}\,\mathrm{I}\sqrt{2}\Big],\,\Big[2-5\,\mathrm{I}\sqrt{2}\Big],\,\frac{199}{200}\Big)$$

$$+49750\,\mathrm{I}\,\mathrm{hypergeom}\Big(\Big[1+\frac{5}{2}\,\mathrm{I}\sqrt{2},\,1+\frac{5}{2}\,\mathrm{I}\sqrt{2}\Big],\,\Big[1+5\,\mathrm{I}\sqrt{2}\Big],\,\frac{199}{200}\Big)\,\mathrm{hypergeom}\Big(\Big[2-\frac{5}{2}\,\mathrm{I}\sqrt{2},\,2-\frac{5}{2}\,\mathrm{I}\sqrt{2}\Big],$$

$$\Big[2-5\,\mathrm{I}\sqrt{2}\Big],\,\frac{199}{200}\Big)\sqrt{2}+30646\,\mathrm{hypergeom}\Big(\Big[2+\frac{5}{2}\,\mathrm{I}\sqrt{2},\,2+\frac{5}{2}\,\mathrm{I}\sqrt{2}\Big],\,\Big[2+5\,\mathrm{I}\sqrt{2}\Big],\,\frac{199}{200}\Big)\,\mathrm{hypergeom}\Big(\Big[1$$

$$-\frac{5}{2}\,\mathrm{I}\sqrt{2},\,1-\frac{5}{2}\,\mathrm{I}\sqrt{2}\Big],\,\Big[1-5\,\mathrm{I}\sqrt{2}\Big],\,\frac{199}{200}\Big)+49750\,\mathrm{I}\,\mathrm{hypergeom}\Big(\Big[2+\frac{5}{2}\,\mathrm{I}\sqrt{2},\,2+\frac{5}{2}\,\mathrm{I}\sqrt{2}\Big],\,\Big[2+5\,\mathrm{I}\sqrt{2}\Big],$$

$$\frac{199}{200}\Big)\sqrt{2}\,\mathrm{hypergeom}\Big(\Big[1-\frac{5}{2}\,\mathrm{I}\sqrt{2},\,1-\frac{5}{2}\,\mathrm{I}\sqrt{2}\Big],\,\Big[1-5\,\mathrm{I}\sqrt{2}\Big],\,\frac{199}{200}\Big)\Big)$$

**Fig. S 4b**

$plot(funcy(x), x = 0..50);$

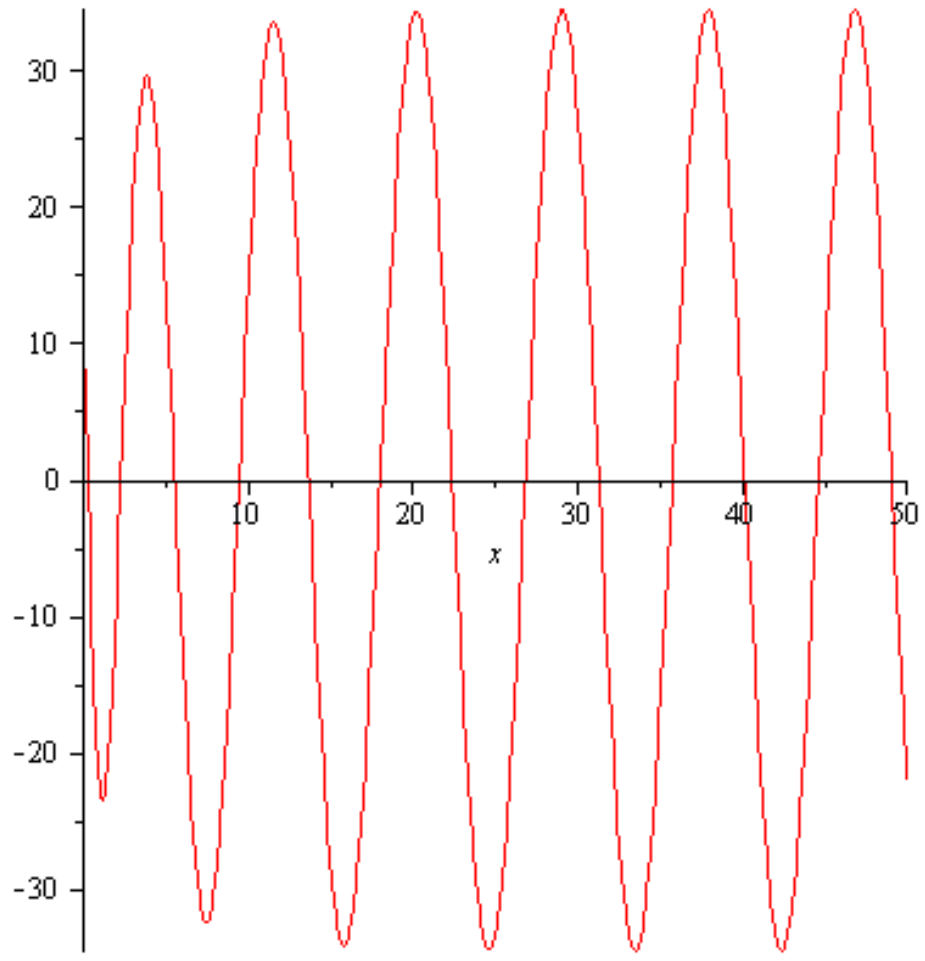

## Supplementary 5: Solutions with learning effects

## Supplementary 5a

$; \ dgl := diff(y(x), x, x) + \exp(x) \cdot y(x) = 0;$

$$\frac{d^2}{dx^2} y(x) + e^x y(x) = 0$$

$dsolve(\{dgl, y(0) = 0, D(y)(0) = 1\}, y(x));$

$$y(x) = \frac{\text{BesselY}(0, 2)\, \text{BesselJ}\left(0, 2\, e^{\frac{1}{2} x}\right)}{-\text{BesselY}(0, 2)\, \text{BesselJ}(1, 2) + \text{BesselY}(1, 2)\, \text{BesselJ}(0, 2)} - \frac{\text{BesselJ}(0, 2)\, \text{BesselY}\left(0, 2\, e^{\frac{1}{2} x}\right)}{-\text{BesselY}(0, 2)\, \text{BesselJ}(1, 2) + \text{BesselY}(1, 2)\, \text{BesselJ}(0, 2)}$$

$funcy := unapply(rhs(dsolve(\{dgl, y(0) = 0, D(y)(0) = 1\}, y(x))), x);$

$$x \rightarrow \frac{\mathrm{BesselY}(0, 2)\, \mathrm{BesselJ}\left(0, 2\, \mathrm{e}^{\frac{1}{2}x}\right)}{-\mathrm{BesselY}(0, 2)\, \mathrm{BesselJ}(1, 2) + \mathrm{BesselY}(1, 2)\, \mathrm{BesselJ}(0, 2)} - \frac{\mathrm{BesselJ}(0, 2)\, \mathrm{BesselY}\left(0, 2\, \mathrm{e}^{\frac{1}{2}x}\right)}{-\mathrm{BesselY}(0, 2)\, \mathrm{BesselJ}(1, 2) + \mathrm{BesselY}(1, 2)\, \mathrm{BesselJ}(0, 2)}$$

**Fig. S 5a**

*plot*( *funcy*(*x*), *x* =−2 ..5);

*plot*( *funcy*(*x*), *x* =−1 ..100);

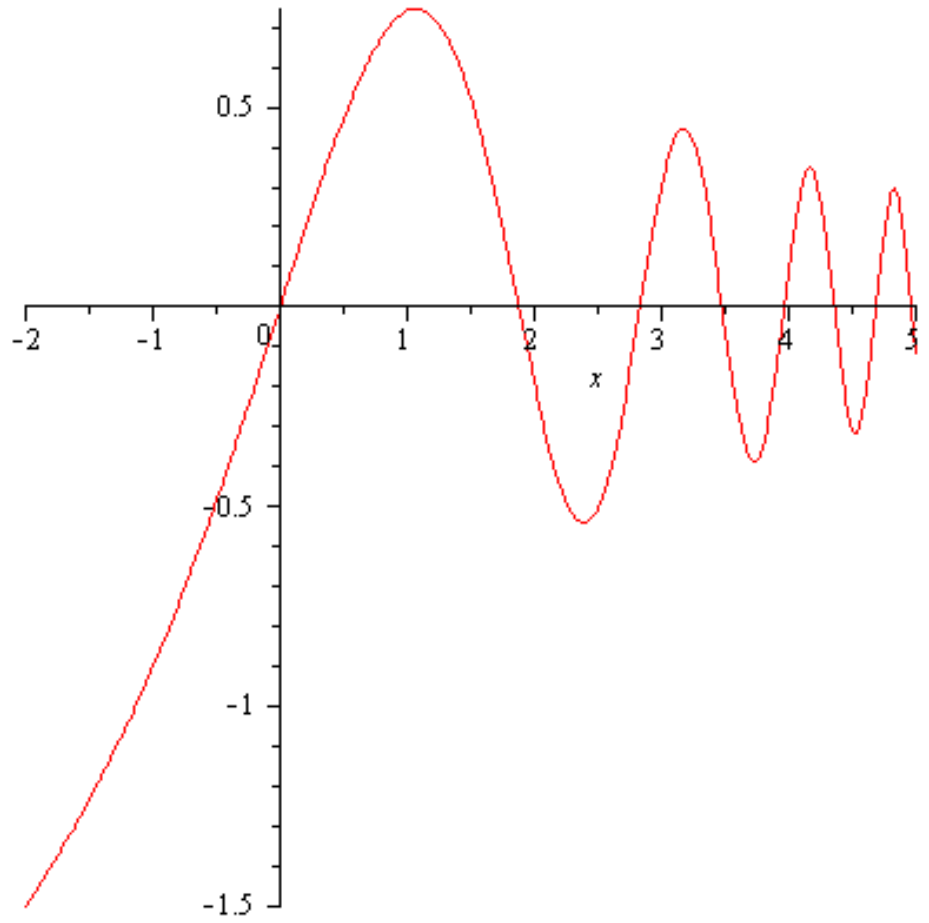

**Supplementary 5b**

*dgl* := *diff*(*y*(*x*), *x*, *x*) + (1 + *x*) ⋅ *y*(*x*) = 0;

$$\frac{d^2}{dx^2}\, y(x) + (1 + x)\, y(x) = 0$$

*dsolve*( {*dgl*, *y*(0) = 0, D(*y*)(0) = 1}, *y*(*x*) );

$$y(x) = \frac{\mathrm{AiryBi}(-1)\,\mathrm{AiryAi}(-1-x)}{-\mathrm{AiryBi}(-1)\,\mathrm{AiryAi}(1,-1)+\mathrm{AiryBi}(1,-1)\,\mathrm{AiryAi}(-1)} - \frac{\mathrm{AiryAi}(-1)\,\mathrm{AiryBi}(-1-x)}{-\mathrm{AiryBi}(-1)\,\mathrm{AiryAi}(1,-1)+\mathrm{AiryBi}(1,-1)\,\mathrm{AiryAi}(-1)}$$

*funcy* := *unapply*(*rhs*(*dsolve*({*dgl*, *y*(0) = 0, D(*y*)(0) = 1}, *y*(*x*))), *x*);

$$x \to \frac{\mathrm{AiryBi}(-1)\,\mathrm{AiryAi}(-1-x)}{-\mathrm{AiryBi}(-1)\,\mathrm{AiryAi}(1,-1)+\mathrm{AiryBi}(1,-1)\,\mathrm{AiryAi}(-1)} - \frac{\mathrm{AiryAi}(-1)\,\mathrm{AiryBi}(-1-x)}{-\mathrm{AiryBi}(-1)\,\mathrm{AiryAi}(1,-1)+\mathrm{AiryBi}(1,-1)\,\mathrm{AiryAi}(-1)}$$

**Fig. S 5b**

*plot*(*funcy*(*x*), *x* =-2 ..5);

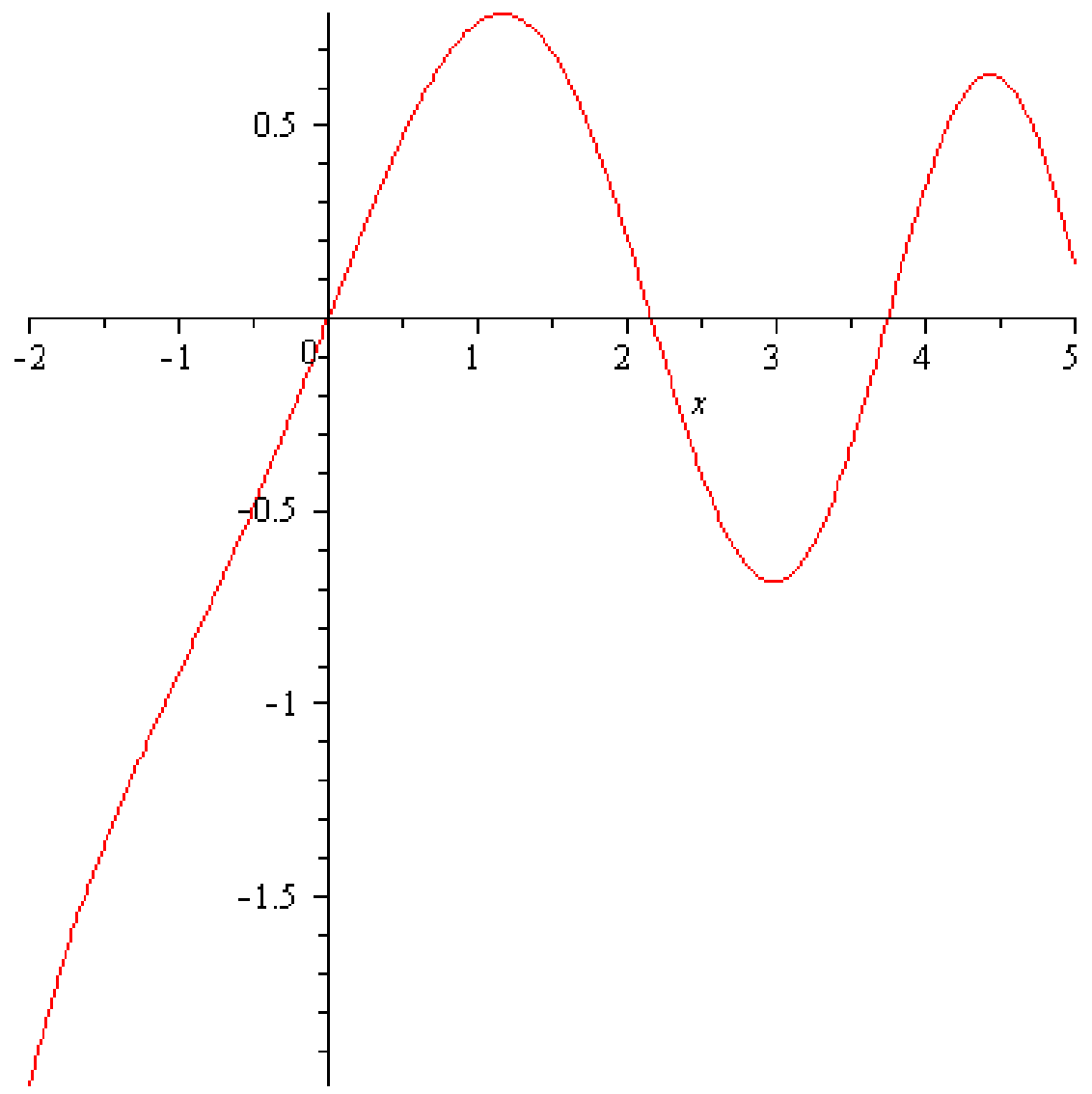

*plot*(*funcy*(*x*), *x* =-1 ..100);

**Supplementary 6**

The combination of a spontaneous with an accumulative reaction leads to a damped oscillatory circuit. See Fig. suppl. 6.

**Fig. S 6**

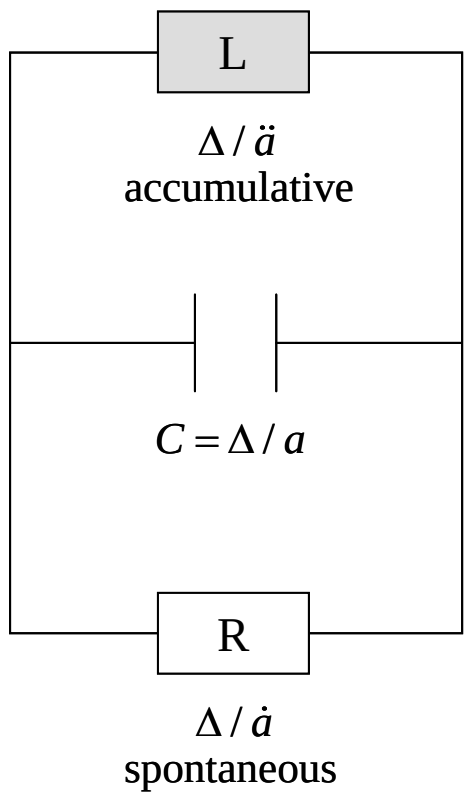

Sociological equivalent circuit that is isomorphic to an electrical LCR circuit. While a LC circuit leads to harmonic oscillation and a RC circuit to exponential relaxation, the LCR circuit shows exponentially damped oscillations, the frequency of which is constant and determined by L, C, R (see textbooks of physics).

# Sociological Cycles:

# The accumulated discrepancy between appearance and reality as driver

Joachim Maier

## Supplementary 1: The fashion problem

This brief consideration deals with intrinsic formation of fashion cycles. (Extrinsic effects such as periodic supply of fashion articles to avoid marked saturation by fashion industry are not taken account of. Even though important in praxi, they are rather trivial in the present context.) We assume two colors blue (B) and red (R). There are two effects of interest. One is the "exclusivity effect" consisting in B when meeting B will change to R (and vice versa). The other is the "imitation or infection effect" consisting in B getting infected by meeting R will change to R (and vice versa). We concentrate on the first effect and set all kinetic constant to unity.

Hence, as for the exclusivity effect

$$B + B \rightarrow R + R$$

and

$$R + R \rightarrow B + B$$

are the important reactions which for simplicity run at the same rates. If we assume simple reaction kinetics both b and r (small characters denote the respective concentrations) will monotonically relax from the initial value to the equilibrium value $\hat{b} = \hat{r} = 1/2$.

If we now introduce a memory such that B changes to R if only B has encountered B n-times and vice versa for R; then we need to distinguish between B's that have had such an encounter for the first, second, … nth time. If e.g. n=2, we need to consider the events

$$
\begin{aligned}
B_0 + B_0 &\rightarrow B_1 + B_1 \\
B_0 + B_1 &\rightarrow B_1 + B_2 \\
B_0 + B_2 &\rightarrow B_1 + R_0 \\
B_1 + B_1 &\rightarrow B_2 + B_2 \\
B_1 + B_2 &\rightarrow B_2 + R_0 \\
B_2 + B_2 &\rightarrow R_0 + R_0
\end{aligned}
$$

and the same for R.

Note that these equations do not satisfy microscopic reversibility and should not be used to describe chemical reactions. Fig. suppl. 1 gives solutions for various n showing that the introduction of a memory leads to damped oscillations (initially 1000 blue and 9000 red particles).

**Fig. S 1**

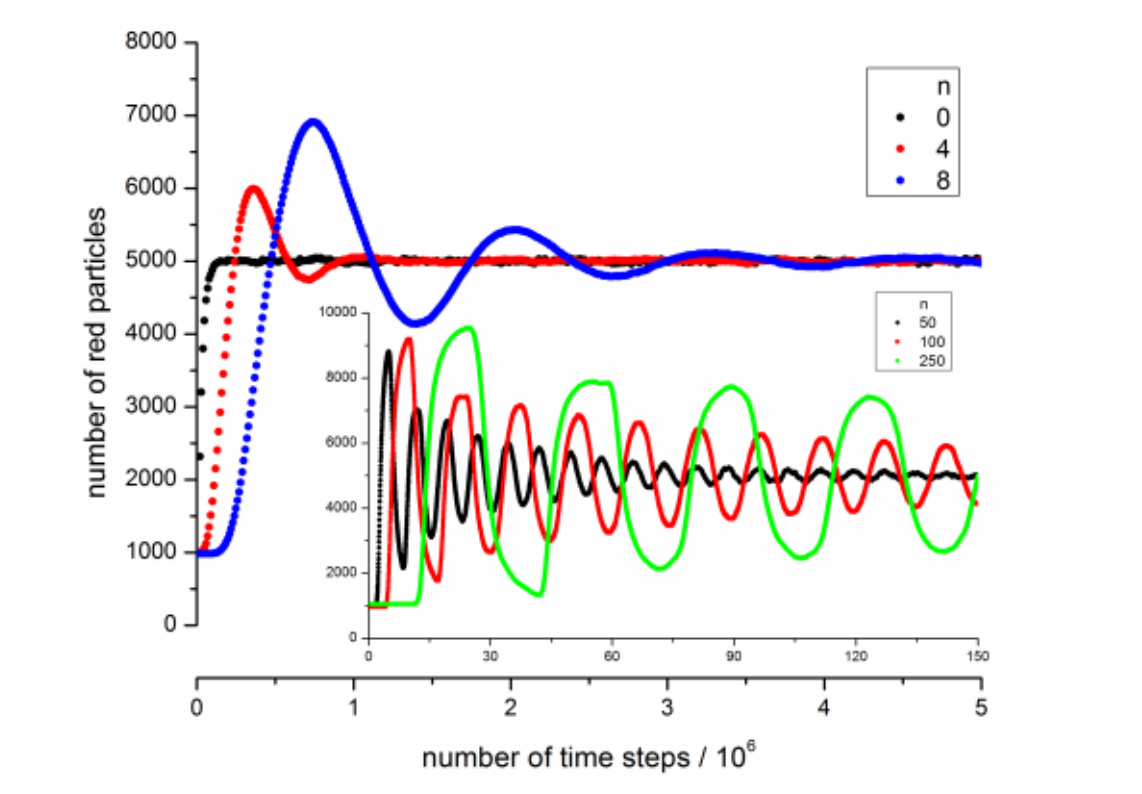

Certainly, the “infection effect”, viz. the tendency of B to change to R on meeting R could be easily implemented as well.

The introduction of different species corresponds to a discrete transition from spontaneous to accumulative behavior. Hence damped oscillations occur the behavior of which depends on the strength of the memory and reaction readiness ( $A_1, ..., A_n$ ; $n$ is related with memory and $\alpha$ ).

A simplified version of the above is obtained if we demand B (or R) to react when meeting a “catalyst” K the concentration of which we can keep constant (see Figure 1, main text).

Then, e.g. for n = 2

$$\begin{aligned} B_0 + K &\rightarrow B_1 + K \\ B_1 + K &\rightarrow B_2 + K \\ B_2 + K &\rightarrow R_0 + K \end{aligned}$$

and the same for R.

Even though these equations are now (since k = const) of first order, the general picture is not very different.

The reaction without memory

$$B + K \rightleftarrows R + K$$

is describable by an exponential function that satisfies the rate law

$$\text{rate} \propto b\text{-}\widehat{b} = -\left(r - \widehat{r}\right)$$

As described in the main text, in order to treat more general situations, we proceed to a description that is not necessarily based on collision events but includes memory by demanding

$$\text{rate} \propto \int \left(b\text{-}\widehat{b}\right) dt.$$

**Supplementary 2:**

*Solutions for* $\ddot{a}(\tau) = a^n(\tau) - a^m(\tau)$ *with selected n, m*:

a) The differential equation $\ddot{a}(\tau) = a^{1/2}(\tau) - a(\tau)$ has the solution

$$\tau = -2\,ArcTan\left[\frac{\left(-2+3\sqrt{a}\right)\sqrt{4a^{3/2}-3a^2}}{\sqrt{3\left(-4+3\sqrt{a}\right)a}}\right]$$

or explicitly for trivial boundary conditions

$$a(\tau) = a_1(\tau) \;@\; a_2(\tau_2)$$

$$a_1 = \frac{4\left(1+3Tan\left[\frac{t}{2}\right]^2+2Tan\left[\frac{t}{2}\right]^4 - 2\sqrt{Tan\left[\frac{t}{2}\right]^2+3Tan\left[\frac{t}{2}\right]^4+3Tan\left[\frac{t}{2}\right]^6+Tan\left[\frac{t}{2}\right]^8}\right)}{9\left(1+2Tan\left[\frac{t}{2}\right]^2+Tan\left[\frac{t}{2}\right]^4\right)}$$

$$a_2 = \frac{4\left(1+3Tan\left[\frac{t}{2}\right]^2+2Tan\left[\frac{t}{2}\right]^4 + 2\sqrt{Tan\left[\frac{t}{2}\right]^2+3Tan\left[\frac{t}{2}\right]^4+3Tan\left[\frac{t}{2}\right]^6+Tan\left[\frac{t}{2}\right]^8}\right)}{9\left(1+2Tan\left[\frac{t}{2}\right]^2+Tan\left[\frac{t}{2}\right]^4\right)}$$

as sketched in Fig. supplementary 2A.

**Fig. S 2A**

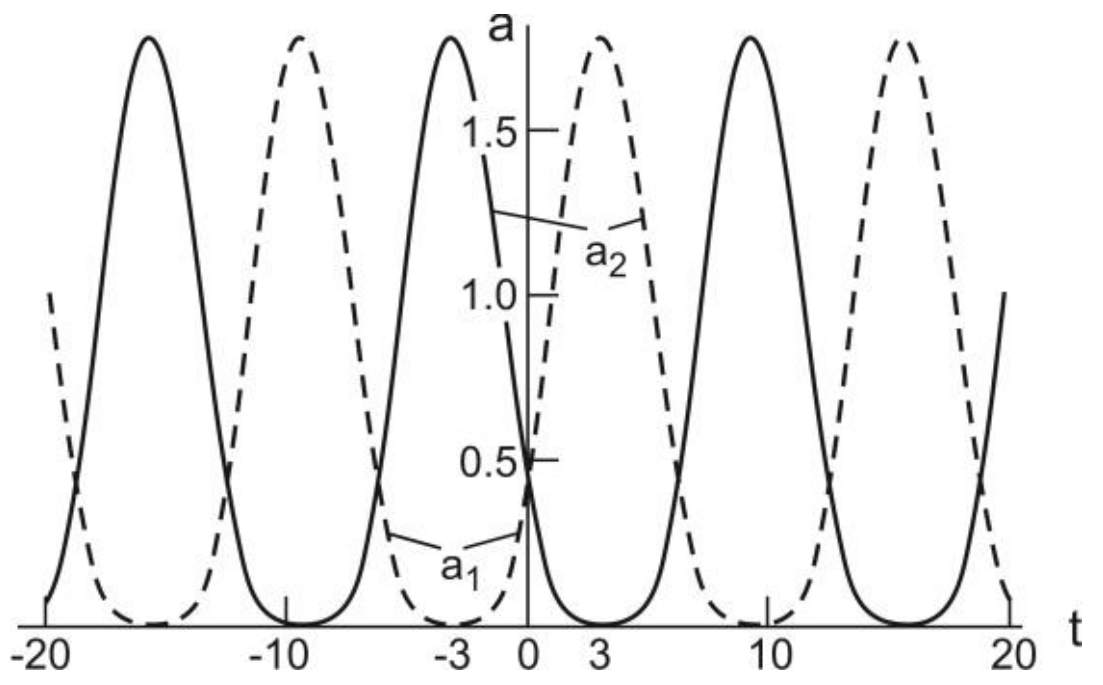

The intercepts with the $\tau$ -axis are odd multiples of π; the intercept with the *a*-axis is 4/9 . The inflection point is at *a* = 1; the maximum is 16/9. The oscillating functions are composites of $a_1$ and $a_2$.

b) The solution for $\ddot{a}(\tau) = a^2(\tau) - a(\tau)$ is for trivial boundary conditions

$$a(\tau) = \frac{3}{2}\left(1 + \left(Tan\left(\frac{\tau}{2}\right)\right)^2\right)$$

The graph is sketched in Fig. suppl. 2B.

**Fig. S 2B**

$$plot\left(\frac{3}{2}\cdot\left(1+\left(\tan\left(\frac{tau}{2}\right)\right)^2\right), tau=-15..15, -10..50\right);$$

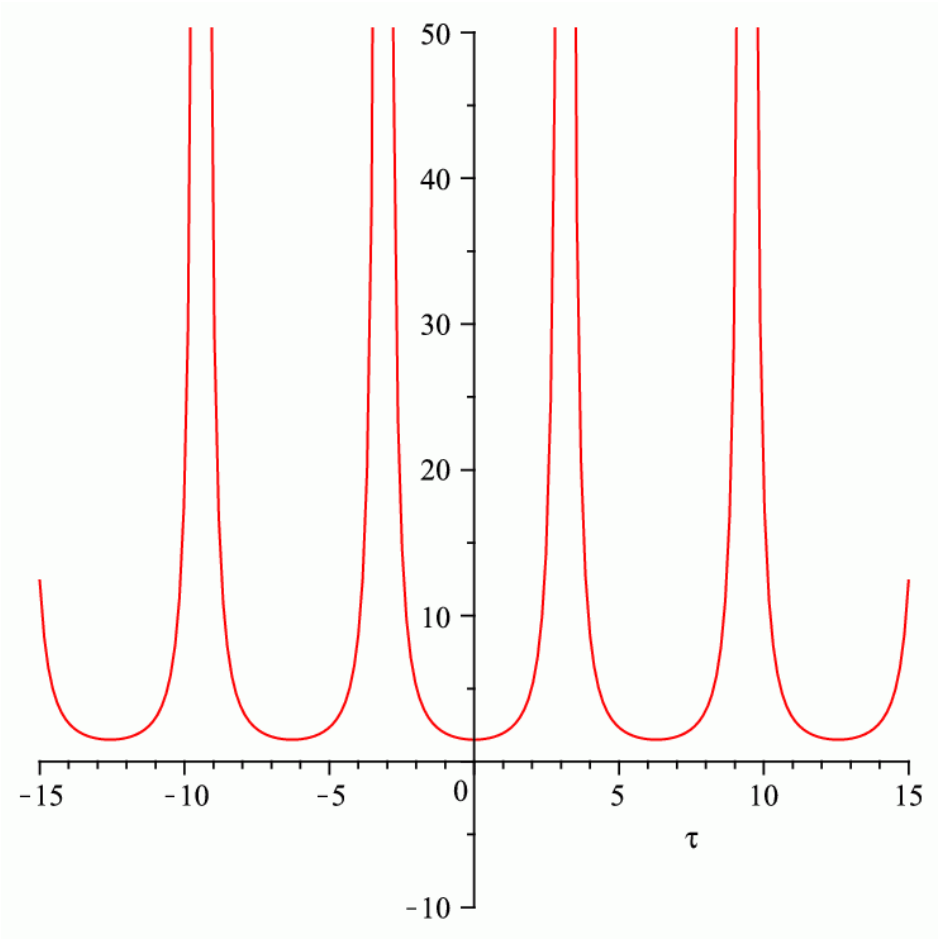

c) The differential equation

$$\ddot{a}(\tau) = a^{3/2}(\tau) - a(\tau)$$

delivers a periodical but non oscillating function

$$a(\tau) = \frac{25}{16}\left[1 + \tan\left[\frac{\tau}{4}\right]^2 + \tan\left[\frac{\tau}{4}\right]^4\right]$$

The $a(\tau)$ curves are given in Fig. supplementary 2 C.

**Fig. S 2C**

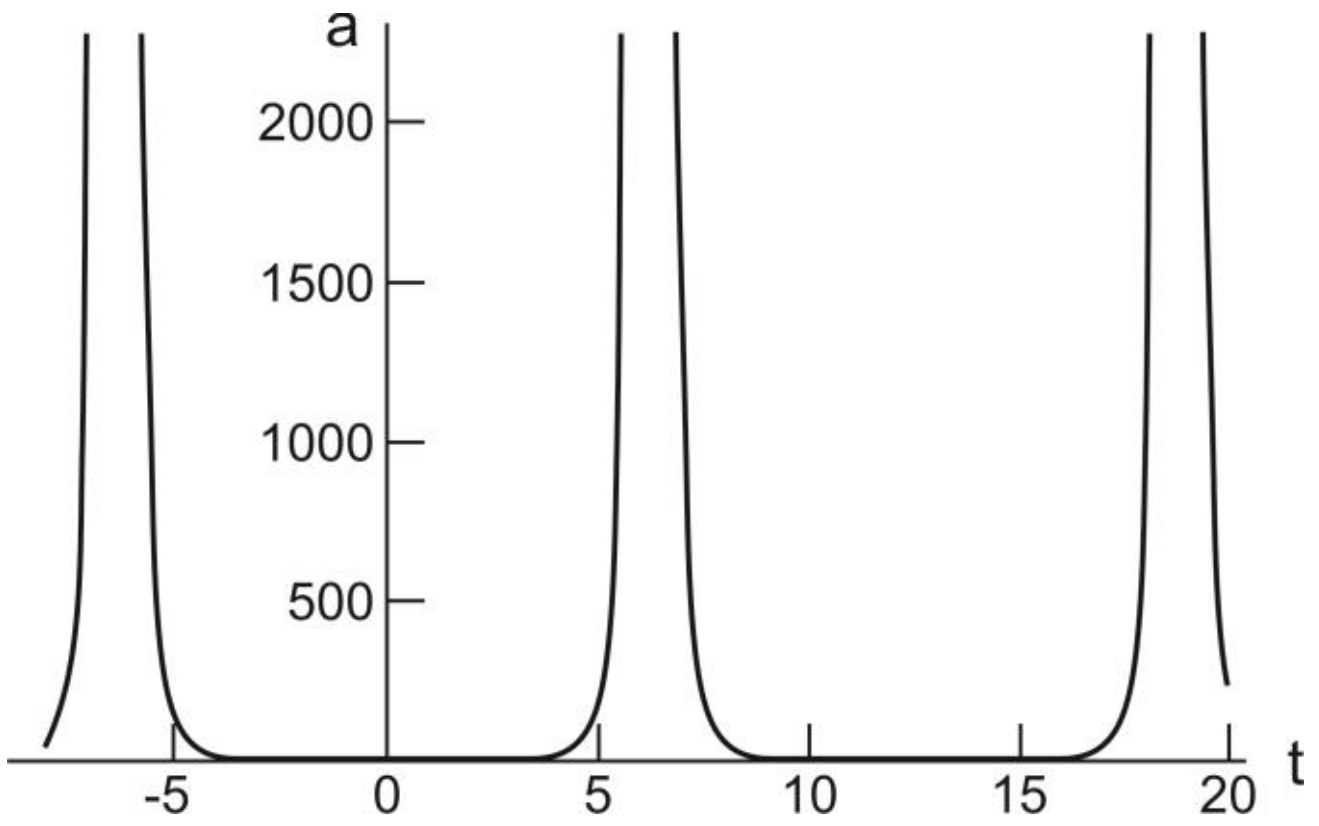

d) It is interesting to compare the solutions for $\ddot{a}(\tau) = a^n(\tau) - a^m(\tau)$ with the spontaneous reaction $\dot{a}(\tau) = a^n(\tau) - a^m(\tau)$. Let us first consider

$$\dot{a}(\tau) = a(\tau) - a^2(\tau)$$

The solution for $a(0) \neq 1$ is

$$a(\tau) = \frac{\dfrac{a(0)}{1 - a(0)}\exp\tau}{1 + \dfrac{a(0)}{1 - a(0)}\exp\tau}$$

For $a < 1$ the first term predominates and $a$ increases. For $a > 1$ the second term predominates and $a$ decreases. Yet, unlike the accumulative response, oscillations do not occur, rather the value $a = 1$ forms a boundary for the solutions. If $a(\tau = 0) \equiv a(0) > 1$, the $a$-function stays always greater than unity, while it stays below unity if $a(0) < 1$. See Fig. suppl. 2D. A similar compartmentation occurs for the solution of

$$\dot{a}(\tau) = a^2(\tau) - a(\tau)$$

which reads for $a(0) \neq 1$

$$a(\tau) = -\frac{\dfrac{a(0)}{a(0)-1}}{\exp\tau - \dfrac{a(0)}{a(0)-1}}$$

**Fig. S 2D**

$$plot\left(\left\{\frac{4}{3}\cdot\frac{\exp(\mathrm{tau})}{1+\left(\frac{4}{3}\right)\cdot\exp(\mathrm{tau})}, -\frac{\exp(\mathrm{tau})}{1-\exp(\mathrm{tau})}\right\}, \mathrm{tau}=-5..5, -5..5\right);$$

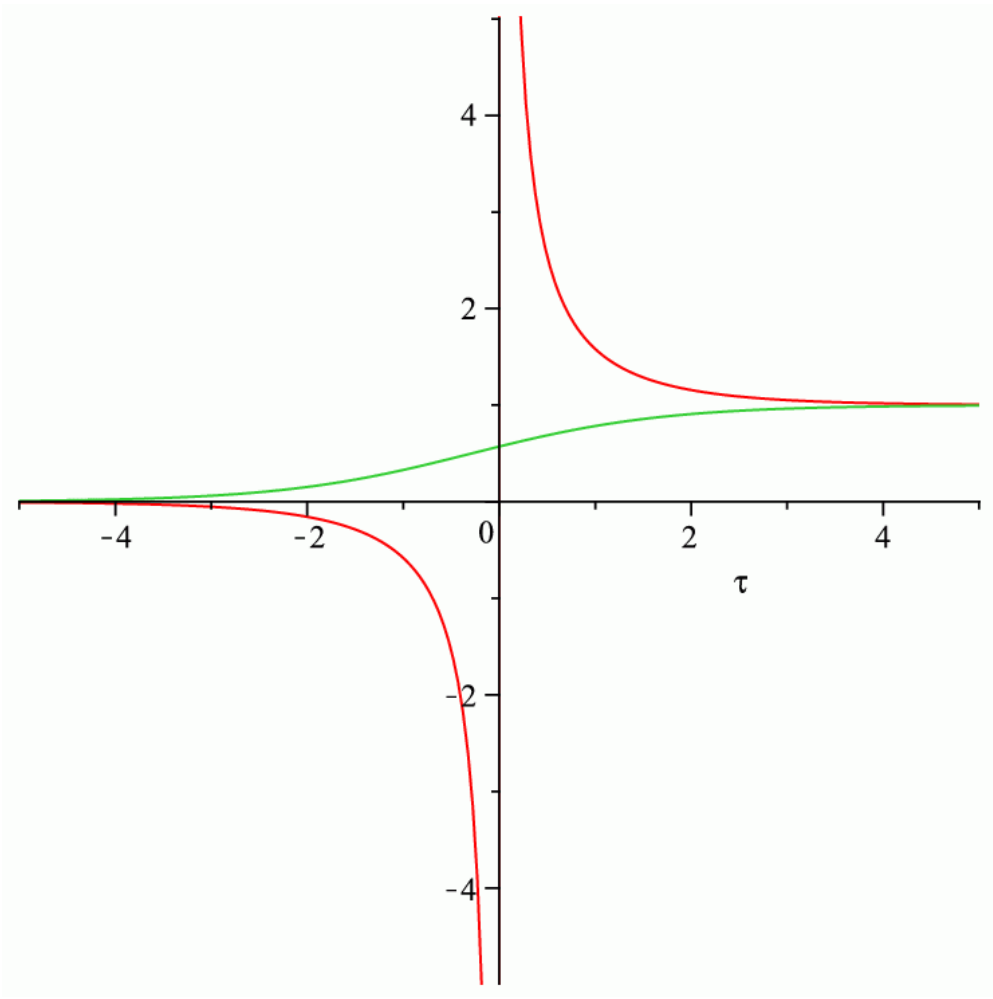

and is displayed in Fig. suppl. 2E, which may be compared with Fig. 2B.

Such a compartmentation is dissolved for an accumulative response.

**Fig. S 2E**

$$plot\left(\left\{-\frac{4}{3}\cdot\left(\exp(tau)-\frac{4}{3}\right)^{-1}, (\exp(tau)\ +1)^{-1}\right\}, tau=-5\,..5, -5\,..5\right);$$

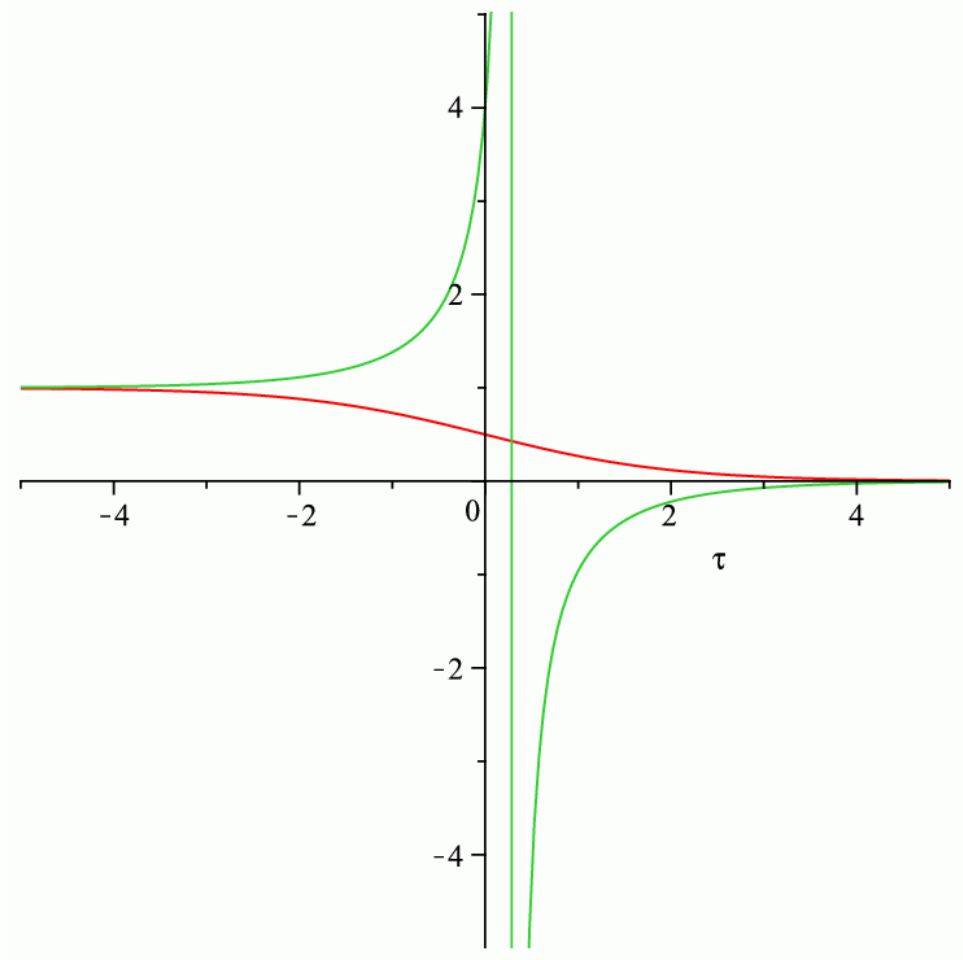

## Supplementary 3: Taboo-phase and breakdown

Let us for simplicity assume direct, spontaneous reaction (rather than an accumulative one) what enables us to adopt the simple formalism of chemical kinetics.

The taboo-phase (or phase of directorship) is determined by the fact that any unwanted action (activity) A must rapidly decay (large reaction constant $\vec{k}_D$) according to

$$A \xrightarrow{\vec{k}_D} \text{Nil}. \qquad (6)$$

Nonetheless there is also a growth mechanism that is enabled wherever a precursor $A^*$ is met ($A^*$ is an activity in a critical pre-stage)

$$A + A^* \xrightarrow{\vec{k}_G} 2A. \qquad (7)$$

The overall rate is $[\mathrm{A}]\left(\vec{k}_G\left[\mathrm{A}^*\right]-\vec{k}_D\right)$. At a high enough number density of $\mathrm{A}^*$ $\left(\text{i.e. of } \left[\mathrm{A}^*\right]\right)$, namely for $\vec{k}_G\left[\mathrm{A}^*\right]>\vec{k}_D$, channel 2 dominates: A grows finally leading to a breakdown of the taboo-phase. The fate of it is determined by the ratio $\frac{\vec{k}_G\left[\mathrm{A}^*\right]}{\vec{k}_D}$. The situation becomes very interesting, if several activities compete with each other and compete about an exhaustible number of $\mathrm{A}^*$. This leads automatically to a selection problem in which the growth (survival) depends on other activities. Incorporation of memory effects to this treatment which follows Ref. (*20*) would lead to analogous solutions for sociological problems.

**Supplementary 4: Solutions including prephases**

**Supplementary 4a**

$dgl := diff(y(x), x, x) + 100\cdot\left(1 - \exp\left(-\frac{x}{10}\right)\right)\cdot y(x) = 0;$

$$\frac{d^2}{dx^2}y(x) + 100\left(1 - e^{-\frac{1}{10}x}\right)y(x) = 0$$

$dsolve(\{dgl, y(0) = -1000, D(y)(0) = 0\}, y(x));$

$$y(x) = \Big(1000\,\big(-\mathrm{BesselK}(1+200\,\mathrm{I}, 200) + \mathrm{I}\,\mathrm{BesselK}(200\,\mathrm{I}, 200)\big)\,\mathrm{BesselI}\big(200\,\mathrm{I}, 200\,\mathrm{e}^{-\frac{1}{20}x}\big)\Big) \Big/ \big(\mathrm{BesselK}(200\,\mathrm{I}, 200)\,\mathrm{BesselI}(1+200\,\mathrm{I}, 200) + \mathrm{BesselK}(1+200\,\mathrm{I}, 200)\,\mathrm{BesselI}(200\,\mathrm{I}, 200)\big) - \Big(1000\,\big(\mathrm{BesselI}(1+200\,\mathrm{I}, 200) + \mathrm{I}\,\mathrm{BesselI}(200\,\mathrm{I}, 200)\big)\,\mathrm{BesselK}\big(200\,\mathrm{I}, 200\,\mathrm{e}^{-\frac{1}{20}x}\big)\Big) \Big/ \big(\mathrm{BesselK}(200\,\mathrm{I}, 200)\,\mathrm{BesselI}(1+200\,\mathrm{I}, 200) + \mathrm{BesselK}(1+200\,\mathrm{I}, 200)\,\mathrm{BesselI}(200\,\mathrm{I}, 200)\big)$$

*funcy* := *unapply*(*rhs*(*dsolve*({*dgl*, *y*(0) = -1000, D(*y*)(0) = 0}, *y*(*x*))), *x*);

$$x \to \Big(1000\,\big(-\mathrm{BesselK}(1+200\,\mathrm{I}, 200) + \mathrm{I}\,\mathrm{BesselK}(200\,\mathrm{I}, 200)\big)\,\mathrm{BesselI}\big(200\,\mathrm{I}, 200\,\mathrm{e}^{-\frac{1}{20}x}\big)\Big) \Big/ \big(\mathrm{BesselK}(200\,\mathrm{I}, 200)\,\mathrm{BesselI}(1+200\,\mathrm{I}, 200) + \mathrm{BesselK}(1+200\,\mathrm{I}, 200)\,\mathrm{BesselI}(200\,\mathrm{I}, 200)\big) - \Big(1000\,\big(\mathrm{BesselI}(1+200\,\mathrm{I}, 200) + \mathrm{I}\,\mathrm{BesselI}(200\,\mathrm{I}, 200)\big)\,\mathrm{BesselK}\big(200\,\mathrm{I}, 200\,\mathrm{e}^{-\frac{1}{20}x}\big)\Big) \Big/ \big(\mathrm{BesselK}(200\,\mathrm{I}, 200)\,\mathrm{BesselI}(1+200\,\mathrm{I}, 200) + \mathrm{BesselK}(1+200\,\mathrm{I}, 200)\,\mathrm{BesselI}(200\,\mathrm{I}, 200)\big)$$

**Fig. S 4a**

*plot*(*funcy*(*x*), *x* = -0.5 ..5);

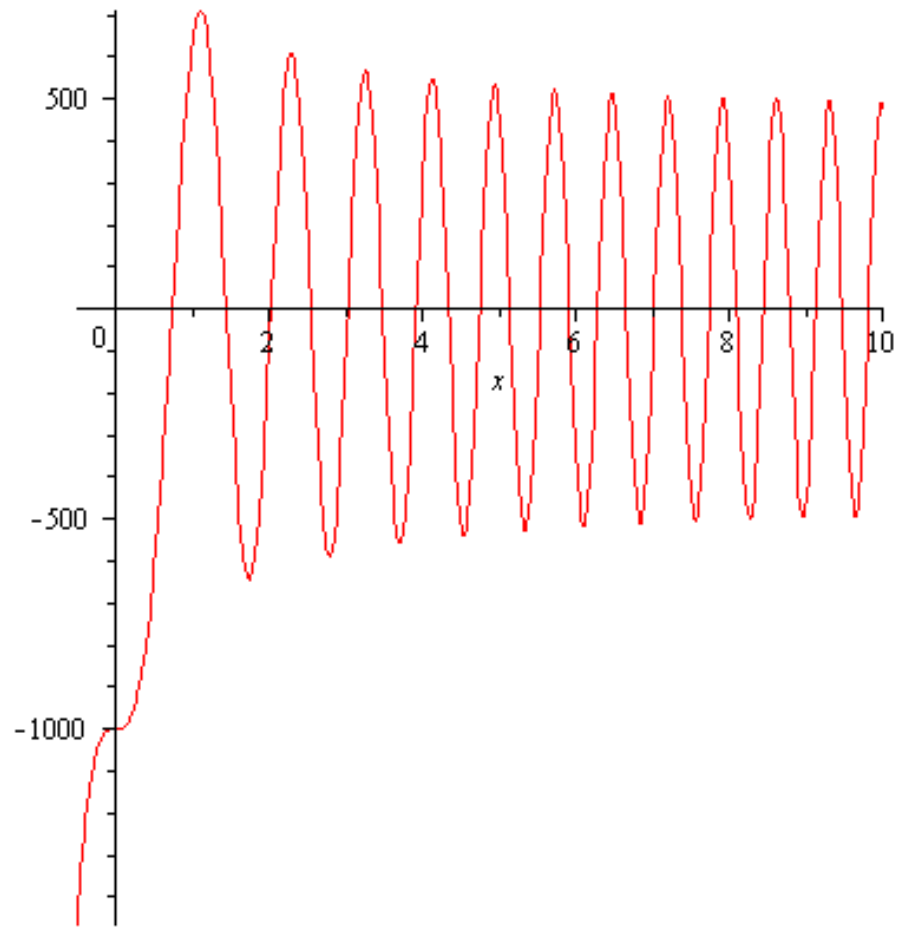

**Supplementary 4b**

$dgl := diff(y(x), x, x) + \left(2 - 1.99 \cdot \exp\left(-\frac{x}{5}\right)\right)^{-1} \cdot y(x) = 0;$

$$\frac{d^2}{dx^2} y(x) + \frac{y(x)}{2 - 1.99\, e^{-\frac{1}{5} x}} = 0 \qquad (1)$$

$dsolve(\{dgl, y(0) = 10, D(y)(0) = 0\}, y(x));$

$$y(x) = -\Bigg(10\Bigg(102000\, I\, \mathrm{hypergeom}\left(\left[1 + \frac{5}{2} I\sqrt{2}, 1 + \frac{5}{2} I\sqrt{2}\right], \left[1 + 5 I\sqrt{2}\right], \frac{199}{200}\right)\sqrt{2} - 8119200\, \mathrm{hypergeom}\left(\left[1 + \frac{5}{2} I\sqrt{2}, 1 + \frac{5}{2} I\sqrt{2}\right], \left[1 + 5 I\sqrt{2}\right], \frac{199}{200}\right) + 30646\, \mathrm{hypergeom}\left(\left[2 + \frac{5}{2} I\sqrt{2}, 2 + \frac{5}{2} I\sqrt{2}\right], \left[2 + 5 I\sqrt{2}\right], \frac{199}{200}\right) + 49750\, I\, \mathrm{hypergeom}\left(\left[2 + \frac{5}{2} I\sqrt{2}, 2 + \frac{5}{2} I\sqrt{2}\right], \left[2 + 5 I\sqrt{2}\right], \frac{199}{200}\right)\sqrt{2}\Bigg)\left(-200\, e^{\frac{1}{2} I\sqrt{2}\, x} + 199\, e^{\frac{1}{2} I\sqrt{2}\, x - \frac{1}{5} x}\right) \mathrm{hypergeom}\left(\left[1 - \frac{5}{2} I\sqrt{2}, 1 - \frac{5}{2} I\sqrt{2}\right], \left[1 - 5 I\sqrt{2}\right], \frac{199}{200}\, e^{-\frac{1}{5} x}\right)\Bigg) \Big/ \Bigg(204000\, I\, \mathrm{hypergeom}\left(\left[1 + \frac{5}{2} I\sqrt{2}, 1 + \frac{5}{2} I\sqrt{2}\right], \left[1 + 5 I\sqrt{2}\right], \frac{199}{200}\right)\sqrt{2}\, \mathrm{hypergeom}\left(\left[1 - \frac{5}{2} I\sqrt{2}, 1 - \frac{5}{2} I\sqrt{2}\right], \left[1 - 5 I\sqrt{2}\right], \frac{199}{200}\right) - 30646\, \mathrm{hypergeom}\left(\left[1 + \frac{5}{2} I\sqrt{2}, 1 + \frac{5}{2} I\sqrt{2}\right], \left[1 + 5 I\sqrt{2}\right], \frac{199}{200}\right) \mathrm{hypergeom}\left(\left[2 - \frac{5}{2} I\sqrt{2}, 2 - \frac{5}{2} I\sqrt{2}\right], \left[2 - 5 I\sqrt{2}\right], \frac{199}{200}\right) + 49750\, I\, \mathrm{hypergeom}\left(\left[1 + \frac{5}{2} I\sqrt{2}, 1\right.\right. \qquad (2)$$

$$+\frac{5}{2}\,\mathrm{I}\sqrt{2}\Big],\ \Big[1+5\,\mathrm{I}\sqrt{2}\Big],\ \frac{199}{200}\Big)\ \mathrm{hypergeom}\Big(\Big[2-\frac{5}{2}\,\mathrm{I}\sqrt{2},\ 2-\frac{5}{2}\,\mathrm{I}\sqrt{2}\Big],\ \Big[2-5\,\mathrm{I}\sqrt{2}\Big],\ \frac{199}{200}\Big)\sqrt{2}$$

$$+30646\ \mathrm{hypergeom}\Big(\Big[2+\frac{5}{2}\,\mathrm{I}\sqrt{2},\ 2+\frac{5}{2}\,\mathrm{I}\sqrt{2}\Big],\ \Big[2+5\,\mathrm{I}\sqrt{2}\Big],\ \frac{199}{200}\Big)\ \mathrm{hypergeom}\Big(\Big[1-\frac{5}{2}\,\mathrm{I}\sqrt{2},\ 1-\frac{5}{2}\,\mathrm{I}\sqrt{2}\Big],\ \Big[1$$

$$-5\,\mathrm{I}\sqrt{2}\Big],\ \frac{199}{200}\Big)+49750\,\mathrm{I}\,\mathrm{hypergeom}\Big(\Big[2+\frac{5}{2}\,\mathrm{I}\sqrt{2},\ 2+\frac{5}{2}\,\mathrm{I}\sqrt{2}\Big],\ \Big[2+5\,\mathrm{I}\sqrt{2}\Big],\ \frac{199}{200}\Big)\sqrt{2}\ \mathrm{hypergeom}\Big(\Big[1$$

$$-\frac{5}{2}\,\mathrm{I}\sqrt{2},\ 1-\frac{5}{2}\,\mathrm{I}\sqrt{2}\Big],\ \Big[1-5\,\mathrm{I}\sqrt{2}\Big],\ \frac{199}{200}\Big)\Big)\Big)-\Big(10\ \Big(102000\,\mathrm{I}\,\mathrm{hypergeom}\Big(\Big[1-\frac{5}{2}\,\mathrm{I}\sqrt{2},\ 1-\frac{5}{2}\,\mathrm{I}\sqrt{2}\Big],$$

$$\Big[1-5\,\mathrm{I}\sqrt{2}\Big],\ \frac{199}{200}\Big)\sqrt{2}+8119200\ \mathrm{hypergeom}\Big(\Big[1-\frac{5}{2}\,\mathrm{I}\sqrt{2},\ 1-\frac{5}{2}\,\mathrm{I}\sqrt{2}\Big],\ \Big[1-5\,\mathrm{I}\sqrt{2}\Big],\ \frac{199}{200}\Big)$$

$$-30646\ \mathrm{hypergeom}\Big(\Big[2-\frac{5}{2}\,\mathrm{I}\sqrt{2},\ 2-\frac{5}{2}\,\mathrm{I}\sqrt{2}\Big],\ \Big[2-5\,\mathrm{I}\sqrt{2}\Big],\ \frac{199}{200}\Big)+49750\,\mathrm{I}\,\mathrm{hypergeom}\Big(\Big[2-\frac{5}{2}\,\mathrm{I}\sqrt{2},\ 2$$

$$-\frac{5}{2}\,\mathrm{I}\sqrt{2}\Big],\ \Big[2-5\,\mathrm{I}\sqrt{2}\Big],\ \frac{199}{200}\Big)\sqrt{2}\Big)\ \Big(-200\,\mathrm{e}^{-\frac{1}{2}\mathrm{I}\sqrt{2}\,x}+199\,\mathrm{e}^{-\frac{1}{2}\mathrm{I}\sqrt{2}\,x-\frac{1}{5}x}\Big)\ \mathrm{hypergeom}\Big(\Big[1+\frac{5}{2}\,\mathrm{I}\sqrt{2},\ 1$$

$$+\frac{5}{2}\,\mathrm{I}\sqrt{2}\Big],\ \Big[1+5\,\mathrm{I}\sqrt{2}\Big],\ \frac{199}{200}\,\mathrm{e}^{-\frac{1}{5}x}\Big)\Big)\Big)\ \Big/\ \Big(204000\,\mathrm{I}\,\mathrm{hypergeom}\Big(\Big[1+\frac{5}{2}\,\mathrm{I}\sqrt{2},\ 1+\frac{5}{2}\,\mathrm{I}\sqrt{2}\Big],\ \Big[1+5\,\mathrm{I}\sqrt{2}\Big],$$

$$\frac{199}{200}\Big)\sqrt{2}\ \mathrm{hypergeom}\Big(\Big[1-\frac{5}{2}\,\mathrm{I}\sqrt{2},\ 1-\frac{5}{2}\,\mathrm{I}\sqrt{2}\Big],\ \Big[1-5\,\mathrm{I}\sqrt{2}\Big],\ \frac{199}{200}\Big)-30646\ \mathrm{hypergeom}\Big(\Big[1+\frac{5}{2}\,\mathrm{I}\sqrt{2},\ 1$$

$$+\frac{5}{2}\,\mathrm{I}\sqrt{2}\Big],\ \Big[1+5\,\mathrm{I}\sqrt{2}\Big],\ \frac{199}{200}\Big)\ \mathrm{hypergeom}\Big(\Big[2-\frac{5}{2}\,\mathrm{I}\sqrt{2},\ 2-\frac{5}{2}\,\mathrm{I}\sqrt{2}\Big],\ \Big[2-5\,\mathrm{I}\sqrt{2}\Big],\ \frac{199}{200}\Big)$$

$$+49750\,\mathrm{I}\,\mathrm{hypergeom}\Big(\Big[1+\frac{5}{2}\,\mathrm{I}\sqrt{2},\ 1+\frac{5}{2}\,\mathrm{I}\sqrt{2}\Big],\ \Big[1+5\,\mathrm{I}\sqrt{2}\Big],\ \frac{199}{200}\Big)\ \mathrm{hypergeom}\Big(\Big[2-\frac{5}{2}\,\mathrm{I}\sqrt{2},\ 2-\frac{5}{2}\,\mathrm{I}\sqrt{2}\Big],$$

$$\Big[2-5\,\mathrm{I}\sqrt{2}\Big],\ \frac{199}{200}\Big)\sqrt{2}+30646\ \mathrm{hypergeom}\Big(\Big[2+\frac{5}{2}\,\mathrm{I}\sqrt{2},\ 2+\frac{5}{2}\,\mathrm{I}\sqrt{2}\Big],\ \Big[2+5\,\mathrm{I}\sqrt{2}\Big],\ \frac{199}{200}\Big)\ \mathrm{hypergeom}\Big(\Big[1$$

$$-\frac{5}{2}\,\mathrm{I}\sqrt{2},\ 1-\frac{5}{2}\,\mathrm{I}\sqrt{2}\Big],\ \Big[1-5\,\mathrm{I}\sqrt{2}\Big],\ \frac{199}{200}\Big)+49750\,\mathrm{I}\,\mathrm{hypergeom}\Big(\Big[2+\frac{5}{2}\,\mathrm{I}\sqrt{2},\ 2+\frac{5}{2}\,\mathrm{I}\sqrt{2}\Big],\ \Big[2+5\,\mathrm{I}\sqrt{2}\Big],$$

$$\frac{199}{200}\Big)\sqrt{2}\ \mathrm{hypergeom}\Big(\Big[1-\frac{5}{2}\,\mathrm{I}\sqrt{2},\ 1-\frac{5}{2}\,\mathrm{I}\sqrt{2}\Big],\ \Big[1-5\,\mathrm{I}\sqrt{2}\Big],\ \frac{199}{200}\Big)\Big)$$

*funcy* := *unapply*(*rhs*(*dsolve*({*dgl*, *y*(0) = 10, D(*y*)(0) = 0}, *y*(*x*))), *x*);

$$x\to-\Big(10\ \Big(102000\,\mathrm{I}\,\mathrm{hypergeom}\Big(\Big[1+\frac{5}{2}\,\mathrm{I}\sqrt{2},\ 1+\frac{5}{2}\,\mathrm{I}\sqrt{2}\Big],\ \Big[1+5\,\mathrm{I}\sqrt{2}\Big],\ \frac{199}{200}\Big)\sqrt{2}-8119200\ \mathrm{hypergeom}\Big(\Big[1 \qquad \textbf{(3)}$$

$$+ \frac{5}{2}\,\mathrm{I}\sqrt{2}, 1 + \frac{5}{2}\,\mathrm{I}\sqrt{2}], [1 + 5\,\mathrm{I}\sqrt{2}], \frac{199}{200}) + 30646\,\mathrm{hypergeom}([2 + \frac{5}{2}\,\mathrm{I}\sqrt{2}, 2 + \frac{5}{2}\,\mathrm{I}\sqrt{2}], [2 + 5\,\mathrm{I}\sqrt{2}],$$

$$\frac{199}{200}) + 49750\,\mathrm{I}\,\mathrm{hypergeom}([2 + \frac{5}{2}\,\mathrm{I}\sqrt{2}, 2 + \frac{5}{2}\,\mathrm{I}\sqrt{2}], [2 + 5\,\mathrm{I}\sqrt{2}], \frac{199}{200})\sqrt{2})\,(-200\,\mathrm{e}^{\frac{1}{2}\mathrm{I}\sqrt{2}\,x}$$

$$+ 199\,\mathrm{e}^{\frac{1}{2}\mathrm{I}\sqrt{2}\,x - \frac{1}{5}x})\,\mathrm{hypergeom}([1 - \frac{5}{2}\,\mathrm{I}\sqrt{2}, 1 - \frac{5}{2}\,\mathrm{I}\sqrt{2}], [1 - 5\,\mathrm{I}\sqrt{2}], \frac{199}{200}\,\mathrm{e}^{-\frac{1}{5}x})))\;\Big/$$

$$(204000\,\mathrm{I}\,\mathrm{hypergeom}([1 + \frac{5}{2}\,\mathrm{I}\sqrt{2}, 1 + \frac{5}{2}\,\mathrm{I}\sqrt{2}], [1 + 5\,\mathrm{I}\sqrt{2}], \frac{199}{200})\sqrt{2}\,\mathrm{hypergeom}([1 - \frac{5}{2}\,\mathrm{I}\sqrt{2}, 1$$

$$- \frac{5}{2}\,\mathrm{I}\sqrt{2}], [1 - 5\,\mathrm{I}\sqrt{2}], \frac{199}{200}) - 30646\,\mathrm{hypergeom}([1 + \frac{5}{2}\,\mathrm{I}\sqrt{2}, 1 + \frac{5}{2}\,\mathrm{I}\sqrt{2}], [1 + 5\,\mathrm{I}\sqrt{2}],$$

$$\frac{199}{200})\,\mathrm{hypergeom}([2 - \frac{5}{2}\,\mathrm{I}\sqrt{2}, 2 - \frac{5}{2}\,\mathrm{I}\sqrt{2}], [2 - 5\,\mathrm{I}\sqrt{2}], \frac{199}{200}) + 49750\,\mathrm{I}\,\mathrm{hypergeom}([1 + \frac{5}{2}\,\mathrm{I}\sqrt{2}, 1$$

$$+ \frac{5}{2}\,\mathrm{I}\sqrt{2}], [1 + 5\,\mathrm{I}\sqrt{2}], \frac{199}{200})\,\mathrm{hypergeom}([2 - \frac{5}{2}\,\mathrm{I}\sqrt{2}, 2 - \frac{5}{2}\,\mathrm{I}\sqrt{2}], [2 - 5\,\mathrm{I}\sqrt{2}], \frac{199}{200})\sqrt{2}$$

$$+ 30646\,\mathrm{hypergeom}([2 + \frac{5}{2}\,\mathrm{I}\sqrt{2}, 2 + \frac{5}{2}\,\mathrm{I}\sqrt{2}], [2 + 5\,\mathrm{I}\sqrt{2}], \frac{199}{200})\,\mathrm{hypergeom}([1 - \frac{5}{2}\,\mathrm{I}\sqrt{2}, 1 - \frac{5}{2}\,\mathrm{I}\sqrt{2}], [1$$

$$- 5\,\mathrm{I}\sqrt{2}], \frac{199}{200}) + 49750\,\mathrm{I}\,\mathrm{hypergeom}([2 + \frac{5}{2}\,\mathrm{I}\sqrt{2}, 2 + \frac{5}{2}\,\mathrm{I}\sqrt{2}], [2 + 5\,\mathrm{I}\sqrt{2}], \frac{199}{200})\sqrt{2}\,\mathrm{hypergeom}([1$$

$$- \frac{5}{2}\,\mathrm{I}\sqrt{2}, 1 - \frac{5}{2}\,\mathrm{I}\sqrt{2}], [1 - 5\,\mathrm{I}\sqrt{2}], \frac{199}{200})) - (10\,(102000\,\mathrm{I}\,\mathrm{hypergeom}([1 - \frac{5}{2}\,\mathrm{I}\sqrt{2}, 1 - \frac{5}{2}\,\mathrm{I}\sqrt{2}],$$

$$[1 - 5\,\mathrm{I}\sqrt{2}], \frac{199}{200})\sqrt{2} + 8119200\,\mathrm{hypergeom}([1 - \frac{5}{2}\,\mathrm{I}\sqrt{2}, 1 - \frac{5}{2}\,\mathrm{I}\sqrt{2}], [1 - 5\,\mathrm{I}\sqrt{2}], \frac{199}{200})$$

$$- 30646\,\mathrm{hypergeom}([2 - \frac{5}{2}\,\mathrm{I}\sqrt{2}, 2 - \frac{5}{2}\,\mathrm{I}\sqrt{2}], [2 - 5\,\mathrm{I}\sqrt{2}], \frac{199}{200}) + 49750\,\mathrm{I}\,\mathrm{hypergeom}([2 - \frac{5}{2}\,\mathrm{I}\sqrt{2}, 2$$

$$- \frac{5}{2}\,\mathrm{I}\sqrt{2}], [2 - 5\,\mathrm{I}\sqrt{2}], \frac{199}{200})\sqrt{2})\,(-200\,\mathrm{e}^{-\frac{1}{2}\mathrm{I}\sqrt{2}\,x} + 199\,\mathrm{e}^{-\frac{1}{2}\mathrm{I}\sqrt{2}\,x - \frac{1}{5}x})\,\mathrm{hypergeom}([1 + \frac{5}{2}\,\mathrm{I}\sqrt{2}, 1$$

$$+ \frac{5}{2}\,\mathrm{I}\sqrt{2}], [1 + 5\,\mathrm{I}\sqrt{2}], \frac{199}{200}\,\mathrm{e}^{-\frac{1}{5}x}))\;\Big/\;(204000\,\mathrm{I}\,\mathrm{hypergeom}([1 + \frac{5}{2}\,\mathrm{I}\sqrt{2}, 1 + \frac{5}{2}\,\mathrm{I}\sqrt{2}], [1 + 5\,\mathrm{I}\sqrt{2}],$$

$$\frac{199}{200})\sqrt{2}\,\mathrm{hypergeom}([1 - \frac{5}{2}\,\mathrm{I}\sqrt{2}, 1 - \frac{5}{2}\,\mathrm{I}\sqrt{2}], [1 - 5\,\mathrm{I}\sqrt{2}], \frac{199}{200}) - 30646\,\mathrm{hypergeom}([1 + \frac{5}{2}\,\mathrm{I}\sqrt{2}, 1$$

$$+ \frac{5}{2}\,\mathrm{I}\sqrt{2}], [1 + 5\,\mathrm{I}\sqrt{2}], \frac{199}{200})\,\mathrm{hypergeom}([2 - \frac{5}{2}\,\mathrm{I}\sqrt{2}, 2 - \frac{5}{2}\,\mathrm{I}\sqrt{2}], [2 - 5\,\mathrm{I}\sqrt{2}], \frac{199}{200})$$

$$+ 49750\,\mathrm{I}\,\mathrm{hypergeom}([1 + \frac{5}{2}\,\mathrm{I}\sqrt{2}, 1 + \frac{5}{2}\,\mathrm{I}\sqrt{2}], [1 + 5\,\mathrm{I}\sqrt{2}], \frac{199}{200})\,\mathrm{hypergeom}([2 - \frac{5}{2}\,\mathrm{I}\sqrt{2}, 2 - \frac{5}{2}\,\mathrm{I}\sqrt{2}],$$

$$[2 - 5\,\mathrm{I}\sqrt{2}], \frac{199}{200})\sqrt{2} + 30646\,\mathrm{hypergeom}([2 + \frac{5}{2}\,\mathrm{I}\sqrt{2}, 2 + \frac{5}{2}\,\mathrm{I}\sqrt{2}], [2 + 5\,\mathrm{I}\sqrt{2}], \frac{199}{200})\,\mathrm{hypergeom}([1$$

$$- \frac{5}{2}\,\mathrm{I}\sqrt{2}, 1 - \frac{5}{2}\,\mathrm{I}\sqrt{2}], [1 - 5\,\mathrm{I}\sqrt{2}], \frac{199}{200}) + 49750\,\mathrm{I}\,\mathrm{hypergeom}([2 + \frac{5}{2}\,\mathrm{I}\sqrt{2}, 2 + \frac{5}{2}\,\mathrm{I}\sqrt{2}], [2 + 5\,\mathrm{I}\sqrt{2}],$$

$$\frac{199}{200})\sqrt{2}\,\mathrm{hypergeom}([1 - \frac{5}{2}\,\mathrm{I}\sqrt{2}, 1 - \frac{5}{2}\,\mathrm{I}\sqrt{2}], [1 - 5\,\mathrm{I}\sqrt{2}], \frac{199}{200}))$$

## Fig. S 4b

*plot*(*funcy*(*x*), *x* = 0 ..50);

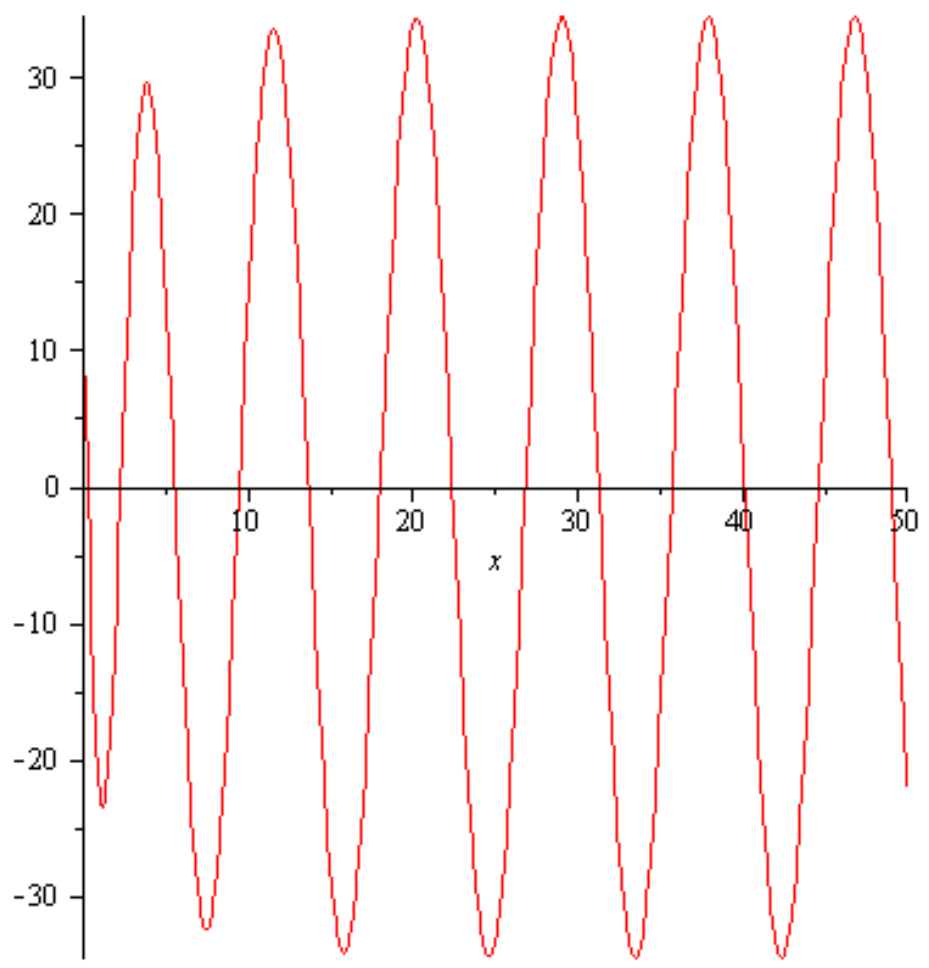

## Supplementary 5: Solutions with learning effects

## Supplementary 5a

; *dgl* := *diff*(*y*(*x*), *x*, *x*) + exp(*x*) ·*y*(*x*) = 0;

$$\frac{d^2}{dx^2}\, y(x) + e^{x}\, y(x) = 0$$

*dsolve*({*dgl*, *y*(0) = 0, D(*y*)(0) = 1}, *y*(*x*));

$$y(x) = \frac{\mathrm{BesselY}(0, 2)\,\mathrm{BesselJ}\left(0, 2\,e^{\frac{1}{2}x}\right)}{-\mathrm{BesselY}(0, 2)\,\mathrm{BesselJ}(1, 2) + \mathrm{BesselY}(1, 2)\,\mathrm{BesselJ}(0, 2)} - \frac{\mathrm{BesselJ}(0, 2)\,\mathrm{BesselY}\left(0, 2\,e^{\frac{1}{2}x}\right)}{-\mathrm{BesselY}(0, 2)\,\mathrm{BesselJ}(1, 2) + \mathrm{BesselY}(1, 2)\,\mathrm{BesselJ}(0, 2)}$$

*funcy* := *unapply*(*rhs*(*dsolve*({*dgl*, *y*(0) = 0, D(*y*)(0) = 1}, *y*(*x*))), *x*);

$$x \to \frac{\mathrm{BesselY}(0,2)\,\mathrm{BesselJ}\left(0, 2\,\mathrm{e}^{\frac{1}{2}x}\right)}{-\mathrm{BesselY}(0,2)\,\mathrm{BesselJ}(1,2) + \mathrm{BesselY}(1,2)\,\mathrm{BesselJ}(0,2)} - \frac{\mathrm{BesselJ}(0,2)\,\mathrm{BesselY}\left(0, 2\,\mathrm{e}^{\frac{1}{2}x}\right)}{-\mathrm{BesselY}(0,2)\,\mathrm{BesselJ}(1,2) + \mathrm{BesselY}(1,2)\,\mathrm{BesselJ}(0,2)}$$

**Fig. S 5a**

*plot*(*funcy*(*x*), *x* = −2 ..5);

*plot*(*funcy*(*x*), *x* = −1 ..100);

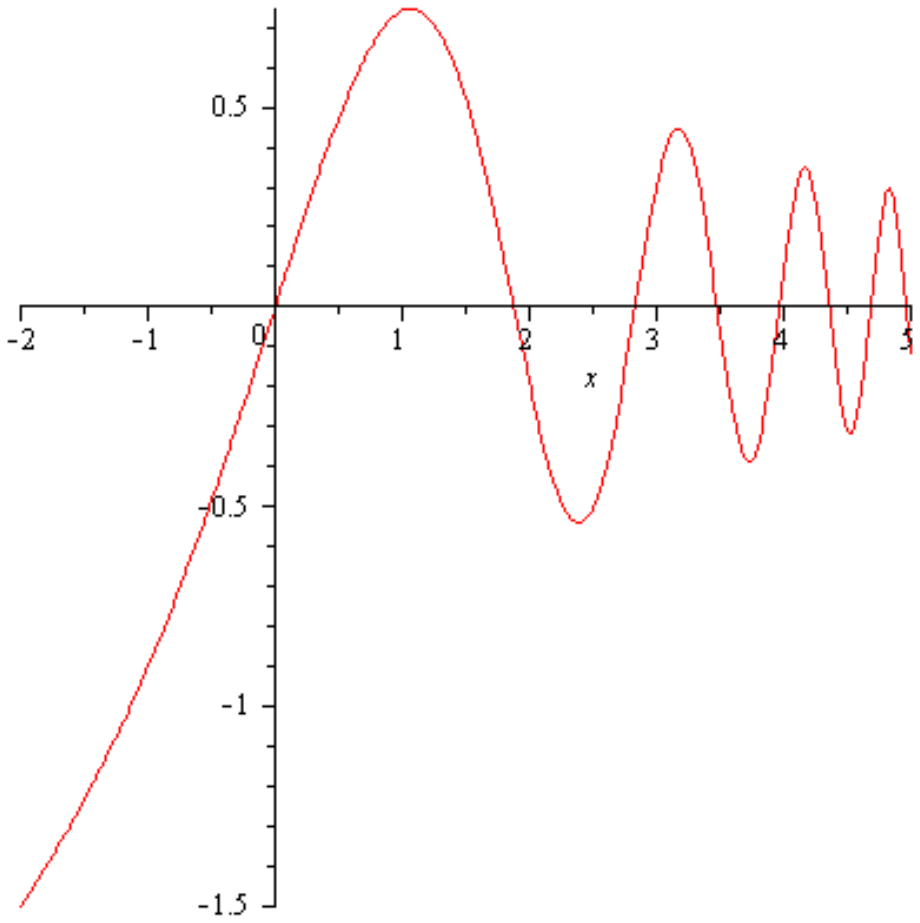

## Supplementary 5b

*dgl* := *diff*(*y*(*x*), *x*, *x*) + (1 + *x*)·*y*(*x*) = 0;

$$\frac{\mathrm{d}^2}{\mathrm{d}x^2}\,y(x) + (1+x)\,y(x) = 0$$

*dsolve*({*dgl*, *y*(0) = 0, D(*y*)(0) = 1}, *y*(*x*));

$$y(x) = \frac{\mathrm{AiryBi}(-1)\,\mathrm{AiryAi}(-1-x)}{-\mathrm{AiryBi}(-1)\,\mathrm{AiryAi}(1,-1) + \mathrm{AiryBi}(1,-1)\,\mathrm{AiryAi}(-1)} - \frac{\mathrm{AiryAi}(-1)\,\mathrm{AiryBi}(-1-x)}{-\mathrm{AiryBi}(-1)\,\mathrm{AiryAi}(1,-1) + \mathrm{AiryBi}(1,-1)\,\mathrm{AiryAi}(-1)}$$

$funcy := unapply(rhs(dsolve(\{dgl, y(0) = 0, \mathrm{D}(y)(0) = 1\}, y(x))), x);$

$$x \rightarrow \frac{\mathrm{AiryBi}(-1)\,\mathrm{AiryAi}(-1-x)}{-\mathrm{AiryBi}(-1)\,\mathrm{AiryAi}(1,-1) + \mathrm{AiryBi}(1,-1)\,\mathrm{AiryAi}(-1)} - \frac{\mathrm{AiryAi}(-1)\,\mathrm{AiryBi}(-1-x)}{-\mathrm{AiryBi}(-1)\,\mathrm{AiryAi}(1,-1) + \mathrm{AiryBi}(1,-1)\,\mathrm{AiryAi}(-1)}$$

**Fig. S 5b**

$plot(funcy(x), x = -2..5);$

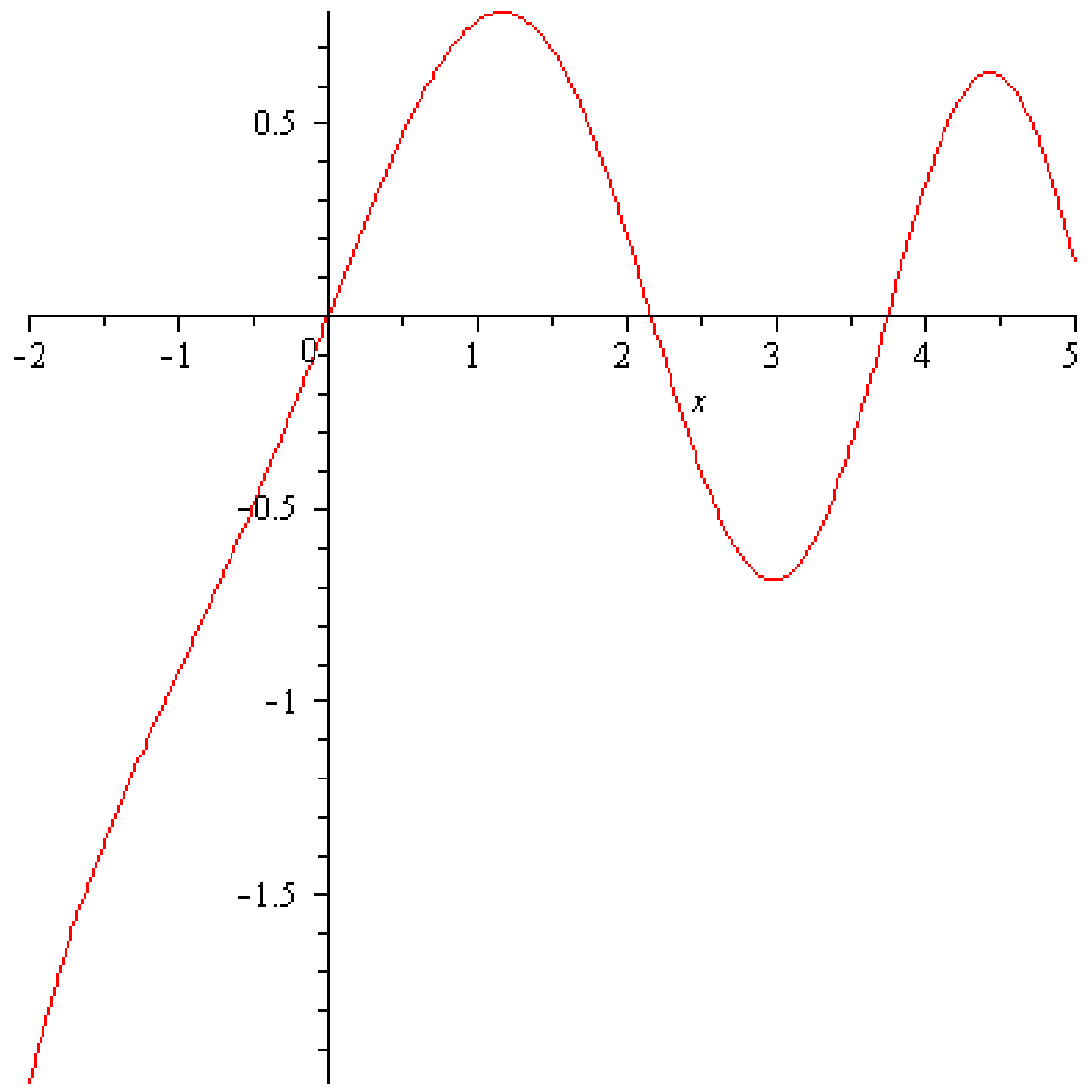

$plot(funcy(x), x = -1..100);$

**Supplementary 6**

The combination of a spontaneous with an accumulative reaction leads to a damped oscillatory circuit. See Fig. suppl. 6.

**Fig. S 6**

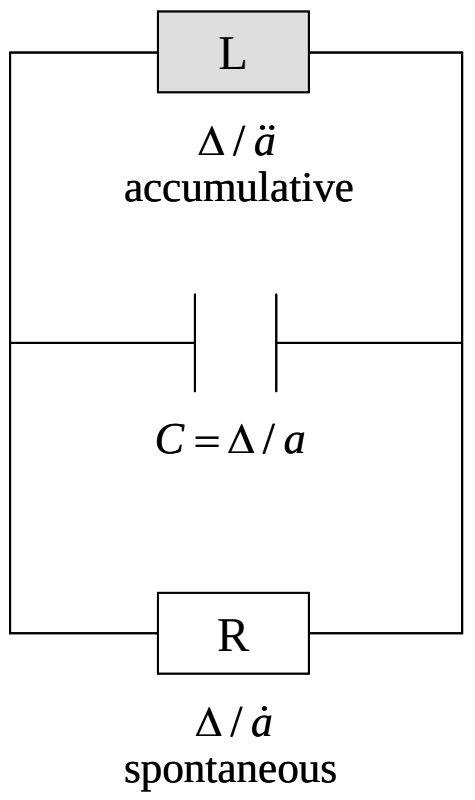

Sociological equivalent circuit that is isomorphic to an electrical LCR circuit. While a LC circuit leads to harmonic oscillation and a RC circuit to exponential relaxation, the LCR circuit shows exponentially damped oscillations, the frequency of which is constant and determined by L, C, R (see textbooks of physics).